\documentclass[prd,preprint,superscriptaddress,preprintnumbers,eqsecnum,showpacs,nofootinbib,nobibnotes]{revtex4-1}
\usepackage{amsfonts,amsmath,mathbbol,amssymb,bm,natbib}
\usepackage[table]{xcolor}
\usepackage{graphicx} 
\usepackage{color}
\usepackage{stackengine}
\usepackage{scalerel}
\usepackage[linkcolor=blue,citecolor=blue,urlcolor=blue,colorlinks=true]{hyperref}
\usepackage{multirow}

\DeclareSymbolFontAlphabet{\amsmathbb}{AMSb}%

\newcommand{\bal}{\begin{align}}
\newcommand{\eal}{\end{align}}

\newcommand{\be}{\begin{equation}}
\newcommand{\ee}{\end{equation}}
\newcommand{\bea}{\begin{eqnarray}}
\newcommand{\eea}{\end{eqnarray}}

\def\s#1{{\scriptscriptstyle #1}}

\def\1eq#1{Eq.~(\ref{#1})}

\def\2eqs#1#2{Eqs.~(\ref{#1}) and~(\ref{#2})}
\def\3eqs#1#2#3{Eqs.~(\ref{#1}),~(\ref{#2}) and~(\ref{#3})}

\def\fig#1{Fig.~\ref{#1}}

\def\ie{{\it i.e.}, }
\def\eg{{\it e.g.}, }

\def\sym{\mathrm{sym}}
\def\asym{\mathrm{asym}}
\newcommand{\bsy}[1]{\bar{#1}}
\newcommand{\Gnp}{\Gamma}
\def\Gone{G_{1}}
\def\Gtwo{G_{2}}
\def\Gthree{G_{3}}

\begin{document}

\title{Novel sum rules for the three-point sector of QCD}

\author{A.~C. Aguilar}
\affiliation{\mbox{University of Campinas - UNICAMP, Institute of Physics ``Gleb Wataghin,''} \\
13083-859 Campinas, S\~{a}o Paulo, Brazil}

\author{M.~N. Ferreira}
\affiliation{\mbox{University of Campinas - UNICAMP, Institute of Physics ``Gleb Wataghin,''} \\
13083-859 Campinas, S\~{a}o Paulo, Brazil}

\author{J. Papavassiliou}
\affiliation{\mbox{Department of Theoretical Physics and IFIC, 
University of Valencia and CSIC},
E-46100, Valencia, Spain}

\begin{abstract}
For special kinematic configurations involving a single momentum scale, 
certain standard relations, originating from the Slavnov-Taylor identities 
of the theory, may be interpreted as ordinary differential equations 
for the  ``kinetic term'' of the gluon propagator.  
The exact solutions of these equations exhibit poles at the origin, which 
are incompatible with the physical answer, known to diverge only
logarithmically; their elimination hinges on the validity
of two integral conditions that we denominate 
 ``asymmetric'' and ``symmetric'' sum rules,
depending on the kinematics employed in their derivation.
The corresponding integrands contain components of the 
three-gluon vertex and the ghost-gluon kernel,  
whose dynamics are constrained when the sum rules are imposed. 
For the numerical treatment we single out 
the asymmetric sum rule, given that its support stems predominantly  
from low and intermediate energy regimes of the defining integral,
which are physically more interesting.
Adopting a combined approach based on 
Schwinger-Dyson equations and lattice simulations, we 
demonstrate how the sum rule clearly favors the suppression of an 
effective form factor entering in the definition of its kernel.
The results of the present work offer an additional vantage point
into the rich and complex structure of the three-point sector of QCD.

\end{abstract}


\maketitle

\section{\label{int}Introduction}

In recent years, the fundamental $n$-point correlation (Green's) functions
of QCD~\cite{Marciano:1977su} have been the subject of systematic scrutiny
both through continuous  methods, such as Schwinger-Dyson equations (SDEs)~\cite{Roberts:1994dr,Alkofer:2000wg,Maris:2003vk,Fischer:2006ub,Roberts:2007ji,Binosi:2009qm,Cloet:2013jya,Binosi:2014aea,Aguilar:2015bud,Binosi:2016rxz,Binosi:2016nme,Huber:2018ned}  and 
functional renormalization group~\cite{Pawlowski:2005xe,Pawlowski:2003hq}, as well as by means of large-volume lattice
simulations~\cite{Sternbeck:2005tk,Ilgenfritz:2006he,Cucchieri:2008fc,Cucchieri:2009zt,Oliveira:2010xc,Maas:2011se,Boucaud:2011ug,Oliveira:2012eh}. In this quest, the original 
intense activity dedicated to the gluon and ghost propagators (two-point sector)~\cite{Aguilar:2006gr, Aguilar:2008xm,Boucaud:2008ky,Fischer:2008uz,
Tissier:2010ts,Campagnari:2010wc,Pennington:2011xs,Aguilar:2011xe,Vandersickel:2012tz,Serreau:2012cg,Fister:2013bh,Kondo:2014sta,Tissier:2017fqf,Corell:2018yil,Cyrol:2017ewj,Gao:2017uox,Cyrol:2018xeq,Kern:2019nzx,Cucchieri:2007md,Cucchieri:2007rg,Bogolubsky:2007ud,Bowman:2007du,Braun:2007bx,Epple:2007ut,Bogolubsky:2009dc,Oliveira:2009eh,Ayala:2012pb,Bicudo:2015rma} 
has been complemented by an in-depth exploration 
of the three-gluon vertex, $\Gamma_{\alpha\mu\nu}$~\cite{Huber:2012kd,Pelaez:2013cpa,Blum:2014gna,Eichmann:2014xya,Vujinovic:2014fza,Cyrol:2016tym,Aguilar:2019jsj,Aguilar:2019uob,Cucchieri:2006tf,Cucchieri:2008qm,Duarte:2016ieu,Athenodorou:2016oyh,Boucaud:2017obn}, the ghost-gluon vertex, 
$\Gamma_{\mu}$~\cite{Boucaud:2011eh,Huber:2012kd,Dudal:2012zx,Aguilar:2013xqa,
Cyrol:2016tym,Mintz:2017qri,Aguilar:2018csq,Cucchieri:2004sq,Ilgenfritz:2006he}, 
and, in part, the auxiliary
ghost-gluon kernel, $H_{\mu\nu}$~\cite{Aguilar:2018csq}. 
This concerted effort has catalyzed a vast array of new theoretical insights on the nonperturbative 
QCD dynamics, and has afforded a tighter grip on a number of complex phenomenological issues
~\cite{Eichmann:2008ef, Cloet:2008re,Meyer:2015eta,Eichmann:2016yit,Sanchis-Alepuz:2017mir,Alkofer:2018guy,Souza:2019ylx,Xu:2019sns,Aguilar:2019teb,Huber:2020ngt}.     

As is well-known, the fundamental Slavnov-Taylor identities (STIs)~\cite{Taylor:1971ff,Slavnov:1972fg} impose
crucial constraints between the two- and three-point sectors of the theory~\cite{Bagan:1989gt, Boucaud:2007va, Aguilar:2010cn,Oliveira:2018fkj, Gracey:2019mix}.
In the present study we offer a novel point of view inspired by these 
profound relations, which, for the special kinematic conditions that
we consider, give rise to two relatively simple sum rules.

The starting point of our considerations are certain special 
projections of $\Gamma_{\alpha\mu\nu}(q,r,p)$, denoted here by $L(q,r,p)$ [see~\1eq{eq:GammaSym_proj1}], 
which have been frequently employed in lattice studies~\cite{Parrinello:1994wd,Alles:1996ka,Parrinello:1997wm,Boucaud:1998bq,Cucchieri:2006tf,Cucchieri:2008qm,Boucaud:2018xup,Aguilar:2019uob}. 
These functions may be evaluated in special kinematic limits, with the final upshot of replacing 
their three momentum scales by a single one. 
In particular, in the so-called ``asymmetric'' and ``symmetric'' 
configurations, the resulting quantities, denoted by $L^\asym(q^2)$ and $L^\sym(s^2)$, respectively,  
have been computed on the lattice, both in quenched~\mbox{\cite{Cucchieri:2006tf,Cucchieri:2008qm,Athenodorou:2016oyh,Duarte:2016ieu,Boucaud:2018xup}},
and unquenched~simulations~\cite{Zafeiropoulos:2019flq,Aguilar:2019uob}.
The basic quantity evaluated in these cases is 
$\langle \widetilde{A}_{\alpha}^{a}(q) \widetilde{A}_{\mu}^{b}(r) \widetilde{A}_{\nu}^{c}(p) \rangle$,
where $\widetilde{A}_{\alpha}^{a}$ are the  SU(3) gauge fields in Fourier space, with  
the average $\langle \cdot \rangle$ denoting functional integration over the gauge space. 

From the continuous standpoint, 
$L^\sym(s^2)$ and $L^\asym(q^2)$ may be written as combinations of the form factors
appearing in the tensorial decomposition of $\Gamma_{\alpha\mu\nu}(q,r,p)$ [see \1eq{eq:3g_sti_structure}];
in particular, $L^\asym(q^2)$ contains only longitudinal  form factors, $X_i(q,r,p)$, 
while $L^\sym(s^2)$ involves both longitudinal and transverse form factors, $Y_i(q,r,p)$~\cite{Ball:1980ax,Davydychev:1996pb,Aguilar:2019kxz}.
The  nonperturbative extension~\cite{Aguilar:2019jsj}  of the Ball-Chiu (BC) procedure~\cite{Ball:1980ax}, in turn,
allows one to relate the $X_i$ with the following quantities: 
{\it (i)} the ``kinetic term'', $J(q^2)$, of the gluon propagator, 
{\it (ii)} the ghost dressing function, $F(q^2)$, and {\it (iii)}  
three of the five form factors, $A_i(q,p,r)$, comprising $H_{\mu\nu}$~\cite{Ball:1980ax,Davydychev:1996pb,Aguilar:2018csq}.
Of course, this STI-based approach leaves the $Y_i$ completely undetermined,
since they form the ``automatically conserved'' part of the three-gluon vertex.
Then, the full implementation of this method gives rise to \2eqs{diffeq_asym}{diffeqsym}~\cite{Aguilar:2019jsj,Aguilar:2019uob}.
Past this point, one introduces theoretical information for the ingredients entering on the
r.h.s. of these equations, thus obtaining definite predictions about $L^\asym(q^2)$ and $L^\sym(s^2)$,
which are subsequently compared with the corresponding lattice results~\cite{Athenodorou:2016oyh,Aguilar:2019uob}.

However, one may reverse this point of view entirely, and consider~\2eqs{diffeq_asym}{diffeqsym} as
relations that furnish $J(q^2)$, once the lattice results for $L^\asym(q^2)$ or $L^\sym(s^2)$ 
have been used as {\it inputs}. 
If this alternative perspective is adopted, it becomes immediately clear that \2eqs{diffeq_asym}{diffeqsym} 
may be viewed as a first order linear differential equations for $J(q^2)$,
whose solution may be written in exact closed form. 

It turns out that the general solution of the differential equation~\eqref{diffeq_asym}
displays a simple pole at the origin, while that of \1eq{diffeqsym} exhibits a double one.
However, it is well known that the {\it physical} $J(q^2)$ {\it does not} possess any type of pole at $q^2=0$; 
instead, as it has been established in a series of works, the {\it massless ghost loop}
entering into the SDE of the $J(q^2)$ forces it to diverge {\it logarithmicaly} as $q^2\to 0$~\cite{Aguilar:2013vaa, Aguilar:2019jsj,Aguilar:2019uob}.

These unphysical poles may be eliminated from the solution for $J(q^2)$ by means of an appropriate expansion
around the origin, provided that certain integral conditions hold exactly.
These conditions, given in \2eqs{condfin_asym}{condfin},
will be referred to as the ``asymmetric'' and ``symmetric'' sum rule, respectively.
At this point, we postpone the determination of $J(q^2)$ from the
corresponding solutions, and focus instead on the
content and potential applications of these sum rules. 

In general, when different sets of ingredients are used as inputs,
the sum rules will be satisfied at a varying degree of accuracy,   
thus providing a quantitative indication on the veracity of the approximations employed for obtaining these ingredients. 
In that sense, the sum rules may be used as a means of discriminating
approximations or truncations schemes, offering hints for their systematic improvement.
Such a possibility, in turn, may be especially useful in the field of SDEs,
where the absence of a concrete expansion parameter complicates the task of  
assigning errors to the results obtained or the simplifications implemented.

 Expanding on the previous point, it should be clear that, 
since the sum rules are deduced from the differential equations for $J(q^2)$, 
the quantities to be probed must be determined from any approach other than the
BC solutions themselves.
For example, restricting ourselves to the asymmetric case, 
one may opt for a purely SDE-based analysis, computing \emph{both} $A_i$
and  $X_i$ from the SDEs of $H_{\mu\nu}$~\cite{Aguilar:2018csq} and $\Gamma_{\alpha\mu\nu}$~\cite{Schleifenbaum:2004id,Huber:2012kd,Aguilar:2013xqa,Huber:2012zj,Blum:2014gna,Eichmann:2014xya,Williams:2015cvx}
respectively, and then plug the $X_i$ into \1eq{eq:asyGamma} to
obtain $L^\asym_{\s {SDE}}(q^2)$. 
Alternatively, one may use a combined approach, deriving the $A_i$  
as before, but using lattice data for $L^\asym(q^2)$~\cite{Athenodorou:2016oyh,Boucaud:2017obn} ; this latter
procedure will be followed in the analysis carried out in Sec.~\ref{num_asym}. 

Since the integrals appearing in \2eqs{condfin_asym}{condfin}
are evaluated within the interval $[0,\mu^2]$, the sum rules
explore the quantities entering in them over a wide range of momenta. 
(we use $\mu=4.3$ GeV throughout).

Note that the symmetric sum rule involves the contributions from
the transverse form factors, $Y_i$, comprising the term $L^\sym_{\s T}(s^2)$,
which enters into the function $\bsy{f}_2(s^2)$ [see \2eqs{splitL}{f1f2}, respectively].
This fact reduces its effectiveness,
at least within the confines of our approach, because the lattice does not furnish
$L^\sym_{\s L}(s^2)$ and $L^\sym_{\s T}(s^2)$ separately, but only their sum.
In addition, as can be seen in \1eq{condfin}, the integrand of the
sum rule contains an additional factor of $t$, with respect to its asymmetric counterpart;
as a result, the support of the ingredients comprising the
kernel is suppressed in the low energy regime, which is the most interesting from the
nonperturbative point of view.
Given the above limitations, for the purposes of this introductory presentation, the numerical
analysis will be restricted to the case of the asymmetric sum rule only.

The article is organized as follows. 
In Sec.~\ref{mainingr} we present a brief summary of
the main ingredients, originating from the two- and three-point sectors of the theory,
that are extensively used in this work. 
Sec.~\ref{special_conf} is dedicated to the detailed derivation of 
\1eq{diffeq_asym}, placing particular emphasis on the origin of the special
function ${\cal W}(q^2)$.
Sec.~\ref{constder} contains the main results of this study. In particular,
after identifying the differential equations and specifying their corresponding solutions,  
we proceed to the detailed derivation of the two sum rules. 
In Sec.~\ref{num_asym} we demonstrate with a concrete example the 
possibilities that  
the asymmetric sum rule offers for constraining one of the ingredients that enter in the
expressions defining the function ${\cal W}(q^2)$. 
In Sec.~\ref{sec:conc} we summarize our results and discuss
future applications of the ideas and techniques presented here. We conclude with two Appendices: 
in the first, we 
implement the transition from the Taylor scheme to the MOM-type scheme 
used in the lattice simulations of $L^\sym(q^2)$; 
in the second, we present the steps necessary for the
one-loop dressed determination of the function ${\cal W}(q^2)$.

\section{\label{mainingr} Brief review of the main theoretical ingredients}

In this section we introduce the necessary notation and
summarize certain basic properties of the two- and three-point correlation functions
entering in this work. 
We emphasize that we restrict ourselves to a ``{\it quenched''} version of QCD,
namely a $SU(3)$ Yang-Mills theory with no dynamical quarks; note, in particular, the {\it absence} of
quark propagators and quark-gluon vertices. 

\subsection{\label{two_sector} Two-point sector: gluon and ghost propagators}

Throughout this article we work in the {\it Landau gauge}, where the 
gluon propagator $\Delta^{ab}_{\mu\nu}(q)=-i\delta^{ab}\Delta_{\mu\nu}(q)$ is given by
the completely transverse form
\begin{align}
\Delta_{\mu\nu}(q) = \Delta(q^2) {\rm P}_{\mu\nu}(q)\,, \qquad {\rm P}_{\mu\nu}(q) = g_{\mu\nu} - \frac{q_\mu q_\nu}{q^2}\,.
\end{align}

The special property of infrared saturation displayed by $\Delta(q^2)$, discussed extensively in the literature cited in the Introduction,
prompts its splitting into two separate components~\cite{Aguilar:2011ux,Binosi:2012sj}, according to (Euclidean space) 
\be
\label{eq:gluon_m_J}
\Delta^{-1}(q^2) = q^2J(q^2) + m^2(q^2)\,,
\ee
where $J(q^2)$ corresponds to the so-called ``kinetic term'' 
while $m^2(q^2)$ represents a momentum-dependent gluon mass scale \mbox{$m^2(0)=\Delta^{-1}(0)$}~
\cite{Smit:1974je,Cornwall:1981zr,Bernard:1981pg,Bernard:1982my,Donoghue:1983fy,Mandula:1987rh,
 Cornwall:1989gv,Wilson:1994fk,Philipsen:2001ip,Aguilar:2002tc,Aguilar:2004sw,Aguilar:2008xm,Binosi:2009qm, Binosi:2012sj,Aguilar:2015bud,Aguilar:2016vin,Binosi:2017rwj}.
In the ultraviolet, the $J(q^2)$ captures the standard perturbative corrections to $\Delta(q^2)$,
while in the infrared it is known to diverge logarithmically~\cite{Aguilar:2013vaa, Aguilar:2019jsj,Aguilar:2019uob}.

In addition, we introduce the ghost propagator, \mbox{ $D^{ab}(q^2)= i \delta^{ab}D(q^2)$},
and the corresponding dressing function, $F(q^2)$, defined by $D(q^2) = F(q^2)/q^2$,
which is known to saturate at a finite value in the deep infrared~\cite{Boucaud:2008ky,Aguilar:2008xm,Dudal:2008sp,Fischer:2008uz,Huber:2018ned}.

\subsection{\label{three_sector}Three-point sector: three-gluon vertex, ghost-gluon kernel and vertex}

Turning to the three-point sector of the theory, we consider:
\begin{enumerate}
\item the three-gluon vertex, $\Gnp^{abc}_{\alpha\mu\nu}(q,r,p) = gf^{abc} \Gnp_{\alpha\mu\nu}(q,r,p)$, 

\item  the ghost-gluon vertex, $\Gamma^{abc}_\mu(q,p,r)=-g f^{abc} \Gamma_\mu(q,p,r)$, 

\item  the ghost-gluon kernel, $H^{abc}_{\nu\mu}(q,p,r) = -gf^{abc}H_{\nu\mu}(q,p,r)$.

\end{enumerate}

\begin{figure}[t]
\begin{center}
\includegraphics[scale=0.6]{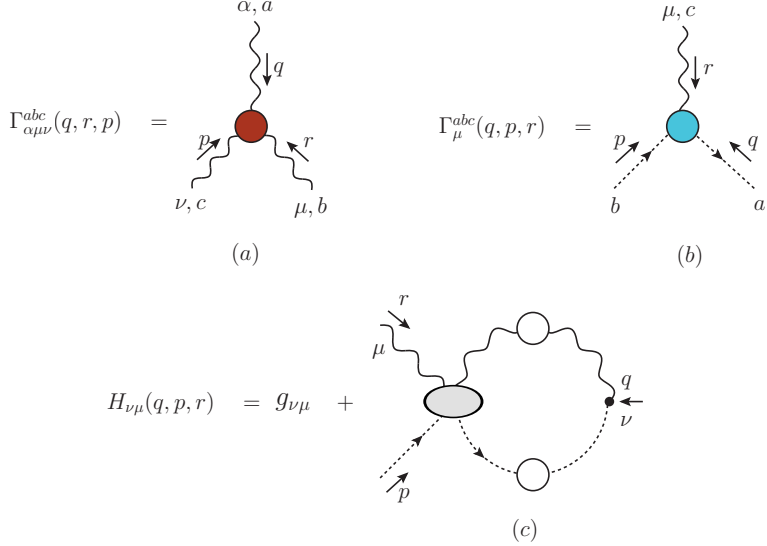}
\end{center}
\vspace{-1.0cm}
\caption{Diagrammatic representations of the  three-gluon vertex, $\Gnp^{abc}_{\alpha\mu\nu}(q,r,p)$, the ghost-gluon vertex,  $\Gamma^{abc}_\mu(q,p,r)$, and the ghost-gluon kernel, $H^{abc}_{\nu\mu}(q,p,r)$  [diagrams  ({\it a}), ({\it b}), and ({\it c}), respectively], with the corresponding  momentum conventions.}
\label{fig:3sector}
\end{figure}

The diagrammatic representations of these three quantities are given in Fig.~\ref{fig:3sector}; 
all momenta are incoming, $q + p + r = 0$.

It is customary to decompose the vertex $\Gnp^{\alpha\mu\nu}(q,r,p)$ in two distinct pieces~\cite{Ball:1980ax,Davydychev:1996pb,Gracey:2014mpa}, according to 
\be
\Gnp^{\alpha\mu\nu}(q,r,p) = \Gamma_{\!\s L}^{\alpha\mu\nu}(q,r,p) + \Gamma_{\!\s T}^{\alpha\mu\nu}(q,r,p)\,,
\label{dec3}
\ee
where the ``longitudinal'' part, $\Gamma_{\!\s L}^{\alpha\mu\nu}(q,r,p)$, saturates the standard STIs
satisfied by the vertex [see \1eq{stig}], 
while the totally ``transverse'' part, $\Gamma_{\!\s T}^{\alpha\mu\nu}(q,r,p)$, is annihilated when 
contracted  by $q_{\alpha}$, $r_{\mu}$, or $p_{\nu}$. The tensorial decomposition of
these two terms reads~\cite{Ball:1980ax,Davydychev:1996pb,Gracey:2014mpa,Aguilar:2019kxz} 
\be
\Gamma_{\!\s L}^{\alpha\mu\nu}(q,r,p) = \sum_{i=1}^{10} X_i(q,r,p) \ell_i^{\alpha\mu\nu} \,,
\qquad \Gamma_{\!\s T}^{\alpha\mu\nu}(q,r,p) = \sum_{i=1}^{4}Y_i(q,r,p)t_i^{\alpha\mu\nu} \,,
\label{eq:3g_sti_structure}
\ee
where the explicit expressions of the basis elements $\ell_i^{\alpha\mu\nu}$ and $t_i^{\alpha\mu\nu}$
are given in~Eqs.~(3.4) and~(3.6) of~\cite{Aguilar:2019jsj}, respectively. 
At tree level,
\be
\Gamma^{(0)}_{\alpha\mu\nu}(q,r,p) = (q-r)_{\nu}g_{\alpha\mu} + (r-p)_{\alpha}g_{\mu\nu} + (p-q)_{\mu}g_{\alpha\nu}\,.
\label{treelevel}
\ee

Regarding $\Gamma_\mu(q,p,r)$ and $H_{\nu\mu}(q,p,r)$, note first that they are related by the STI 
\mbox{$q^\nu H_{\nu\mu}(q,p,r) = \Gamma_\mu(q,p,r)$}. Their respective tensorial decompositions are given by~\mbox{\cite{Ball:1980ax,Davydychev:1996pb,Aguilar:2018csq}}
\bea
\Gamma_\mu (q,p,r) &=& q_\mu B_1(q,p,r) + r_\mu B_2(q,p,r)\,,
\label{Gammamu}
\\
H_{\nu\mu}(q,p,r) &=& g_{\mu\nu}A_1 + q_\mu q_\nu A_2 + r_\mu r_\nu A_3 + q_\mu r_\nu A_4 + r_\mu q_\nu A_5\,,
\label{eq:H}
\eea
where the argument $(q,p,r)$ of the $A_i$ has been suppressed for compactness.
At tree-level, $B_1^{(0)}=1$ and $B_2^{(0)}=0$, while 
$A^{(0)}_1 = 1$ and $A^{(0)}_i = 0$, for $i=2,\ldots,5$. In addition, it is convenient to
introduce the short-hand notation
\be
A_{\s{d}}(q,p,r) := A_3(q,p,r)-A_4(q,p,r) \,.
\label{A34}
\ee

Finally, of central importance for the ensuing analysis is the STI 
\be
q^\alpha\Gnp_{\alpha\mu\nu}(q,r,p) = F(q^2)[p^2J(p^2) {\rm P}^{\alpha}_\nu(p)H_{\alpha\mu}(p,q,r) - r^2J(r^2){\rm P}^{\alpha}_\mu(r)H_{\alpha\nu}(r,q,p)] \,,
\label{stig}
\ee
and its cyclic permutations~\cite{Ball:1980ax,Aguilar:2019jsj},
relating the two- and three-point sectors of the theory. The validity of this set of STIs,
in conjunction with 
the nonperturbative generalization of the standard BC construction~\cite{Ball:1980ax},
allows one to express the longitudinal form factors,  
$X_i$, in terms of $F(q^2)$, $J(q^2)$, $A_1(q,r,p)$, $A_3(q,r,p)$, and $A_{\s{d}}(q,r,p)$~\cite{Aguilar:2019jsj}, but leaves  
the transverse ones, $Y_i$, completely undetermined.

\section{\label{special_conf} The asymmetric configuration}

Of central interest in what follows is the quantity
\begin{equation}
L(q,r,p) = \frac{W^{\alpha\mu\nu}(q,r,p){\rm P}_{\alpha'\alpha}(q){\rm P}_{\mu'\mu}(r){\rm P}_{\nu'\nu}(p)\Gnp^{\alpha'\mu'\nu'}(q,r,p)}
{W^{\alpha\mu\nu}(q,r,p)W_{\alpha\mu\nu}(q,r,p)}\,, 
\label{eq:GammaSym_proj1}
\end{equation}
employed in numerous lattice simulations of the three-gluon vertex~\cite{Athenodorou:2016oyh,Aguilar:2019uob}.
$W^{\alpha\mu\nu}(q,r,p)$ represents specific tensors, which project out 
particular components of the $\Gnp^{\alpha'\mu'\nu'}(q,r,p)$, evaluated in special kinematic limits~\cite{Athenodorou:2016oyh,Aguilar:2019uob}.

In the present work we will focus on the \emph{asymmetric limit},  corresponding to the kinematic configuration 
\be 
p \to 0\,, \qquad r = - q \,.  
\label{defasym}
\ee

Then, an appropriate choice for the $W^{\alpha\mu\nu}$ [see Eq.~(2.22) of~\cite{Aguilar:2019uob}] 
give rise to the special version of $L(q,r,p)$, denoted by $L^\asym(q^2)$,   
which have been computed on the lattice, in quenched~\cite{Cucchieri:2006tf,Cucchieri:2008qm,Athenodorou:2016oyh,Duarte:2016ieu,Boucaud:2018xup}  as well as unquenched~\cite{Aguilar:2019uob} simulations.

As has been shown in~\cite{Aguilar:2019jsj,Aguilar:2019uob}, $L^\asym(q^2)$ is given by
\be   
\label{eq:asyGamma}
L^\asym(q^2)=  X_1(q,-q, 0) - q^2 X_3(q,-q, 0)\, ;
\ee
since it contains no transverse form factors, $Y_i$, $L^\asym(q^2)$ is {\it fully}  
determined from the BC solution for $X_1$ and $X_3$.
The $A_i$ entering in this solution appear in  {\it two different}
kinematic configurations, $A_i(q,-q,0)$ and $A_i(q,0,-q)$, corresponding to 
the {\it soft gluon} and {\it soft ghost} limits, respectively;
we will employ the short-hand notation 
\be
A_i(q^2) := A_i(q,-q,0)\,, \qquad  \widetilde{A}_i(q^2) := A_i(q,0,-q) \,.
\label{Ai_Aitilde}
\ee

Note that, by virtue of Taylor's theorem~\cite{Taylor:1971ff}, in the Landau gauge, $\widetilde{A}_1(q^2)$ is a \mbox{$q^2$-independent} constant [see Appendix~\ref{renor} for details];
therefore, in what follows we simply set \mbox{$\widetilde{A}_1(q^2) \to \widetilde{A}_1$}.

The case of $X_1(q,-q,0)$ can be read off directly from~\cite{Aguilar:2019jsj}; specifically, after conversion to Euclidean space,
\be 
X_1(q,-q,0) = F(q^2)J(q^2) A_1(q^2) \,. 
\label{X1_asym_Ai}
\ee

The form factor $X_3(q,r,p)$ is given by (Minkowski space) 
\be 
X_3(q,r,p) = \frac{F(p^2)}{q^2 - r^2} \Big[ J(q^2) \Gone(q,r,p) - J(r^2) \Gone(r,q,p) \Big] \,, 
\label{X1_X3_general_spher}
\ee
where
\be 
\Gone(q,r,p) := A_1(q,p,r) + (q\cdot p) A_{\s{d}}(q,p,r) \,.
\label{Xnum_def}
\ee

Evidently, due to the  vanishing of the denominator, the $p = 0$ limit requires an appropriate expansion.
Using that \mbox{$q^2 - r^2 = - 2 \, q\cdot p + {\mathcal O}(p^2)$}, and expanding $J(r^2)$ in the numerator,
{\it i.e.} \mbox{$J(r^2) = J(q^2) + 2\, q\cdot p \, J'(q^2)$}, we obtain 
\be 
X_3(q,-q,0) = F(0) \left[ J'(q^2) \widetilde{A}_1 - J(q^2) \left( \widetilde{A}_{\s{d}}(q^2) + \Gthree(q^2) \right) \right] \,, 
\label{X3_p0_step1}
\ee
\be 
\Gthree(q^2) := \lim_{ p\to 0 } \frac{\Gtwo(q,r,p)}{2\, ( q\cdot p ) } \,, \qquad \Gtwo(q,r,p) := A_1(q,p,r) - A_1(r,p,q) \,.
\label{DA_def}
\ee

In order to evaluate $\Gthree(q^2)$ further, note that 
$A_1(q,p,r)$ is obtained from $H_{\nu\mu}(q,p,r)$ through the projection $A_1(q,p,r) = {\cal T}_{\! 1}^{\mu\nu}(q,r) H_{\nu\mu}(q,p,r)$, 
where the projector ${\cal T}_{\! 1}^{\mu\nu}(q,r)$ satisfies  
\be 
   {\cal T}_{\! 1}^{\mu\nu}(r,q) = {\cal T}_{\! 1}^{\mu\nu}(q,r)\,; \quad {\cal T}_{\! 1 \,\, \mu}^{\,\, \mu}(q,r) = 1\,; \quad {\cal T}_{\! 1}^{\mu\nu}(q,r) \, t_\mu = 0 \,; \quad {\cal T}_{\! 1}^{\mu\nu}(q,-q) = \frac{{\rm P}_{\mu\nu}(q)}{3} \,,
\label{T1cont}
\ee
with $t=q,r,p$ [see Eq.~(3.8) of~\cite{Aguilar:2018csq}]. 
Moreover, in the Landau gauge, all quantum corrections to $H_{\nu\mu}(q,p,r)$ are proportional to $p$, such that~\cite{Ibanez:2012zk}
\be 
H_{\nu\mu}(q,p,r) = g_{\mu\nu} + p^\rho K_{\nu\mu\rho}(q,p,r) \,.
 \label{H_tay}
\ee
Hence, the $\Gtwo(q,r,p)$ of \1eq{DA_def} may be cast in the form 
\be 
\Gtwo(q,r,p) =p^\rho \,  {\cal T}_{\! 1}^{\mu\nu}(q,r) \, \left[ K_{\nu\mu\rho}(q,p,r) - K_{\nu\mu\rho}(r,p,q) \right ] \,.
\label{DA2}
\ee
Evidently, the limit $p=0$ may be taken directly in the expression in square brackets, yielding
\be 
\Gtwo(q,r,p) = p^\rho \,  {\cal T}_{\! 1}^{\mu\nu}(q,-q) \, \left[ K_{\nu\mu\rho}(q,0,-q) - K_{\nu\mu\rho}(- q,0,q) \right ] + {\cal O}(p^2) \,.
 \label{DA3}
\ee
Next, $K_{\nu\mu\rho}(q,0,-q)$ may be written as 
\be 
K_{\nu\mu\rho}(q,0,-q) = - \frac{{\cal W}(q^2)}{q^2} g_{\mu\nu}q_\rho + \cdots \,, 
\label{HKtens}
\ee
where the ellipses denote terms proportional to $g_{\mu\rho}q_\nu$,  $g_{\rho\nu}q_\mu$, and $q_\mu q_\nu q_\rho$,
which do not contribute to $\Gtwo(q,r,p)$, by virtue of \1eq{T1cont}. Hence, we have
\be 
\Gtwo(q,r,p) = - 2 \, ( q \cdot p )  \frac{ {\cal W}(q^2)}{q^2} + {\cal O}(p^2) \,, 
\label{DHK}
\ee
which, upon substitution into \1eq{DA3}, leads to $\Gthree(q^2) =  - {\cal W}(q^2)/q^2$.
Using the last expression into \1eq{X3_p0_step1}, and passing to Euclidean space ($q^2 \to - q^2$), we finally get
\be 
X_3(q,-q,0) = - F(0) J'(q^2) \widetilde{A}_1(q^2) - F(0) J(q^2) \left( \widetilde{A}_{\s{d}}(q^2) + \frac{{\cal W}(q^2)}{q^2} \right) \,. \label{X3_asym_Ai}
\ee

Substituting \2eqs{X1_asym_Ai}{X3_asym_Ai} into \1eq{eq:asyGamma} leads immediately to
\be 
L^\asym(q^2) = F(q^2) J(q^2) A_1( q^2 ) + q^2 F(0) \left[ J'(q^2) \widetilde{A}_1 + \left(\widetilde{A}_{\s{d}}(q^2) + \frac{{\cal W}(q^2)}{q^2} \right) J(q^2) \right]  \,;
\label{diffeq_asym_noSTI}
\ee
the detailed derivation  of the function ${\cal W}(q^2)$
is given in Appendix~\ref{a1tan_deriv} [see in particular~\eqref{Wproj}].

The above equation may be simplified considerably by resorting
to the exact relation
\be 
\frac{F(r^2)[ A_1(q,r,p) - p^2 A_3(q,r,p) - ( q \cdot p ) A_4(q,r,p) ]}{F(p^2)[ A_1(q,p,r) - r^2 A_3(q,p,r) - ( q \cdot r ) A_4(q,p,r) ]} = 1 \,, 
\label{STIconst}
\ee
which is a direct consequence of a fundamental STI satisfied by $H_{\nu\mu}$~\cite{Binosi:2011wi,Aguilar:2018csq}. In particular, setting  in it $p = 0$ and  $r = - q$,  we obtain 
\be 
q^2 \widetilde{A}_{\s{d}}(q^2) = \widetilde{A}_1 - \frac{ F(q^2) }{ F(0) } A_1(q^2) \,.
\label{A43const}
\ee

The substitution of the above result into \1eq{diffeq_asym_noSTI} eliminates all dependence on $A_1(q^2)$, $\widetilde{A}_3(q^2)$, and $\widetilde{A}_4(q^2)$,
yielding the compact result
\be 
L^\asym(q^2) = F(0)\left[ J(q^2)\left( \widetilde{A}_1 + {\cal W}( q^2 ) \right) + q^2 J'(q^2) \widetilde{A}_1 \right] \,.
\label{diffeq_asym}
\ee

It is clear from \1eq{diffeq_asym} that the logarithmic divergence displayed by $J(q^2)$ in the deep infrared
is transferred to $L^\asym(q^2)$. In particular, from \1eq{a1tan_sde} follows that \mbox{${\cal W}(0) = 0$} 
[for more details, see the end of Appendix~\ref{a1tan_deriv}]; 
moreover, $\widetilde{A}_1 = Z^\asym_1$ [see Eq.~\eqref{z2_def} in the Appendix~\ref{renor}], while the term $\lim\limits_{q^2\to 0} q^2 J^{\prime}(q^2)$ is {\it subleading}, contributing a finite constant. 
Thus, the {\it leading} contribution of \1eq{diffeq_asym} is given by 
\be
\lim\limits_{q^2\to 0} L^\asym(q^2)  =  Z^\asym_1 F(0) \lim\limits_{q^2\to 0} J(q^2) \,,
\label{condor_asym}
\ee
relating the rates of divergence of $L^\asym(q^2)$ and $J(q^2)$ at the origin.

\section{\label{constder} Derivation of the sum rules}

It is clear that the STI-derived relation given in \1eq{diffeq_asym}
may be regarded as a first order linear differential equation for $J(q^2)$, whose solution allows one  
to express $J(q^2)$ in terms of all other functions. It turns out that this particular point of view,
when appropriately explored, leads to two novel constraints, whose detailed derivation is the
focal point of this section.

\subsection{\label{constasym} Asymmetric sum rule}

Let us consider \1eq{diffeq_asym}, set $x = q^2$, and define
\be
{f_1}(x) = 1 + \frac{{\cal W}( x )}{\widetilde{A}_1} \,, \qquad {f_2}(x) = \frac{L^\asym(x)}{F(0) \widetilde{A}_1} \,.
\label{f1_f2_asym} 
\ee
Then, \1eq{diffeq_asym} may be 
cast in the ``canonical'' form of a linear differential equation 
\be 
J'(x) + P(x)J(x) = Q(x) \,,
\label{odeas}
\ee
with
\be 
P(x) = \frac{{f_1}(x)}{x} \,, \qquad Q(x) = \frac{{f_2}(x)}{x} \,. 
\label{PandQ}
\ee

Therefore, the solution of Eq.~\eqref{odeas} reads~\cite{Arfken:7thed}
\be
J(x) = \frac{1}{\lambda(x)} \left[\lambda(\mu^2) J(\mu^2) + \int_{\mu^2}^x dt \lambda(t) Q(t)\right] \,,
\label{solgen_asym1}
\ee
where $\lambda(x)$ is the ``integrating factor'', given by
\be
\lambda(x) = \exp \left[\int\!\! dx\, P(x) \right] \,,
\label{intfac}
\ee
and $\mu^2$ is the point where the initial condition is chosen.

At this point, it is natural to opt for an initial condition dictated by the physics, identifying $\mu^2$
with the subtraction point where $J(x)$ has been renormalized.
Specifically, in the momentum subtraction scheme (MOM) usually employed,  
we have that $J(\mu^2) = 1$, so that \1eq{solgen_asym1} becomes
\be
J(x) = \frac{1}{\lambda(x)} \left[\lambda(\mu^2)  + \int_{\mu^2}^x\!\! dt \,\lambda(t) \,Q(t)\right]\,.
\label{solgen_asym2}
\ee
The particularity of this solution originates from the presence of the $x$
in the denominator of $P(x)$, which, in general, introduces in the answer a pole divergence at $x=0$.

To see this property in its most rudimentary manifestation, set into \1eq{f1_f2_asym}  
${\cal W}(x) = 0$ (tree-level value), so that ${f}_1(x) = 1$, while ${f}_2(x)$ is kept arbitrary.
Then, the integrating factor becomes simply 
\be
\lambda(x) = x \,,
\ee
and the solution reads
\be
J(x) = \frac{1}{x} \left[\mu^2  + \int_{\mu^2}^x\!\! dt\, {f}_2(t)\right] \,.
\label{solgen_asym3}
\ee

Now, let us suppose that we know from independent considerations that
$J(x)$ does {\it not} diverge as a pole at $x=0$, but rather
as a logarithm~\cite{Aguilar:2013vaa, Aguilar:2019jsj,Aguilar:2019uob}.
Then, the question that arises naturally is how to reconcile this information with the form of \1eq{solgen_asym3}.

Perhaps the most direct approach for answering this question is to consider the Taylor expansion around $x=0$ 
of the expression in square brackets on the r.h.s. of~\1eq{solgen_asym3}. Specifically, one has 
\be
\mu^2  + \int_{\mu^2}^x\!\! dt\, {f}_2(t) = a_0 + a_1 x + {\cal O}(x^2) \,, 
\label{TayExp}
\ee
with
\be
a_0 = \mu^2 + \int_{\mu^2}^0 \!\!dt \, {f}_2(t)\,, \qquad a_1 =  {f}_2(0) \,.
\ee
Clearly, in order for the solution not to possess a pole at the origin, we must have
$a_0 =0$.
This condition amounts to the integral constraint   
\be
\int^{\mu^2}_0 \!\!\!dt \, {f}_2(t) = \mu^2 \,,
\label{cond1}
\ee
which must be obeyed by the function ${f}_2(t)$ within the 
interval of integration $[0,\mu^2]$.

If the above condition is satisfied, then the solution of the differential equation at $x=0$ yields  
\mbox{$J(0) = {f}_2(0)$}, which is none other than the leading term of \1eq{condor_asym}.

Let us emphasize that the constraint~\eqref{cond1} does not hinge on the
specifics of the behavior of $J(x)$, other than the fact that it does not
display an $1/x$ divergence as $x\to 0$. For example, regardless of whether 
$J(x)$ displays near the origin the logarithmic behavior advocated in the literature, or
goes simply to a constant, the corresponding ${f}_2(t)$ is bound to satisfy \1eq{cond1}.

Having fixed the ideas, 
let us next consider the complete case, where the function 
${f}_1(x)$ retains its full structure. Noting that ${\cal W}(0) = 0$, we have that ${f_1}(0) = 1$; it is then convenient to define 
\be
{u}(x) := {f_1}(x) - 1 = \frac{{\cal W}(x)}{\widetilde{A}_1} \,,
\label{bar_u}
\ee
with ${u}(0)=0$.  Then, \1eq{intfac} yields
\be
\lambda(x) = x {\sigma}(x) \,, \qquad {\sigma}(x) := \exp\left[\int \! dx\, {u}(x)/x \right] \,.
\label{theg_asym}
\ee
As is clear from \1eq{H_tay}, $H_{\nu\mu}$, $K_{\nu\mu\rho}$, 
and their respective form factors, $\widetilde{A}_1$ and ${\cal W}(q^2)$, are all renormalized by the 
the same (finite) constant, $Z^\asym_1$, which drops out when forming the
ratio that defines ${u}(x)$.
Consequently, both ${u}(x)$ and $\sigma(x)$
are ``renormalization group invariant''($\mu$-independent) quantities.

With the definitions introduced in \1eq{theg_asym}, the solution in \1eq{solgen_asym2} becomes 
\be
J(x) = \frac{1}{ x \, {\sigma}(x) }\left[ \mu^2 {\sigma}(\mu^2) + \int_{\mu^2}^x \!\!dt \, {\sigma}( t ) \, {f_2}(t) \right] \,, 
\label{solgen_asym}
\ee
and the Taylor expansion around $x = 0$ can be carried through as in \1eq{TayExp}.
At this point, the requirement of eliminating from the solution the pole at $x=0$ imposes the constraint
\be 
\int_0^{\mu^2}\!\!\! dt \, {\sigma}( t ) \, {f_2}(t) = \mu^2 {\sigma}(\mu^2) \,,
\label{condfin_asym} 
\ee
which is a central result of this work, to be referred to as the ``\emph{asymmetric sum rule}''. 

Next, assuming the constraint of \1eq{condfin_asym} to be satisfied, the asymptotic behavior of $J(q^2)$ is obtained from the first nonvanishing term of the Taylor expansion, yielding \emph{exactly} the leading relation reported in \1eq{condor_asym}.

\subsection{\label{altder} Alternative derivation}

Let us return to \1eq{diffeq_asym}, but, instead of solving it for $J(x)$, use 
\1eq{eq:gluon_m_J} to set $J(x) = [\Delta^{-1}(x) -m^2(x)]/x$, and convert it into a differential equation for $m^2(x)$, treating   
$\Delta(x)$ as an input known from lattice simulations. Then, straightforward algebra yields
\be
[m^2(x)]^{\prime} +  P_m(x) \, m^2(x) = Q_m(x) \,,
\label{odem}
\ee
with 
\be
P_m(x) = \frac{{\cal W}(x)}{x\widetilde{A}_1} \,,\qquad  Q_m(x) = - {f_2}(x) +
[\Delta^{-1}(x)]^{\prime} + \frac{{\cal W}(x)}{x\widetilde{A}_1}\Delta^{-1}(x)\,.
\ee
It is clear at this point that the integrating factor, $\lambda_m(x)$, is given by 
\mbox{$\lambda_m(x) = {\sigma}(x)$},  where ${\sigma}(x)$ is  defined in Eq.~\eqref{theg_asym},
and that  \mbox{${\cal W}(x)/[ x\widetilde{A}_1 ] = {\sigma}^{\prime}(x)/{\sigma}(x)$}, which allows us to write $Q_m(x)$ in the form
\be
Q_m(x) = - {f_2}(x) + \frac{[\Delta^{-1}(x) {\sigma}(x)]^{\prime}}{{\sigma}(x)}\,.
\ee
Then, choosing the boundary condition at the origin, namely $m^2(0)= \Delta^{-1}(0)$,
and using that ${\sigma}(0)=1$, the solution for $m^2(x)$ is given by
\bea 
m^2(x) &=& \frac{1}{{\sigma}(x)}\left[ \Delta^{-1}(0) + \int_{0}^x \!\!dt \, {\sigma}(t) \, Q_m(t) \right]  \label{solm_asym}
\nonumber\\
&=& \frac{1}{{\sigma}(x)}\left[ \Delta^{-1}(0) + \left[\Delta^{-1}(t) {\sigma}(t)\right ]^x_0 - \int_{0}^x \!\!dt \, {\sigma}(t) {f_2}(t) \right ]
\nonumber\\
&=& \Delta^{-1}(x) -  \frac{1}{{\sigma}(x)} \int_{0}^x \!\!dt \,{\sigma}(t){f_2}(t)\,.
\label{meq}
\eea
The direct comparison between the last line of \1eq{meq} and \1eq{eq:gluon_m_J} prompts immediately the 
identification 
\be
J(x) = \frac{1}{x {\sigma}(x)} \int_{0}^x \!\!dt \, {\sigma}(t) {f_2}(t)\,.
\label{Jagain}
\ee
If at this point we impose the normalization condition $J(\mu^2) = 1$ at the level of \1eq{Jagain},
we recover directly the constraint~\eqref{condfin_asym}.

At first sight, it would seem that the constraint of \1eq{condfin_asym} has now been obtained without the key assumption
that $J(x)$ should not have a pole. Note, however, that this property is implicitly assumed as soon as 
one uses the relation $m^2(0)= \Delta^{-1}(0)$; indeed, if $J(x)$ had a pole, then $xJ(x)$ would furnish a constant
at the origin, and one could not attribute the value $\Delta^{-1}(0)$ exclusively to $m^2(0)$.

We emphasize that \2eqs{Jagain}{solgen_asym} are completely equivalent, as can be seen
immediately by using $\int_{\mu^2}^x = \int_{0}^{\mu^2} - \int_{0}^x$ in \1eq{solgen_asym} and subsequently employing
\1eq{condfin_asym}. 
Furthermore, note that once the solution
for $J(x)$ has been cast in the form of \1eq{Jagain}, its pole 
may be explicitly removed by means of the simple change of variables $t = x y$.

\subsection{\label{symapp} Symmetric sum rule}

Another special version of the $L(q,r,p)$ defined in \1eq{eq:GammaSym_proj1}, to be denoted by 
$L^\sym(s^2)$,  
corresponds to the \emph{totally symmetric limit}~\cite{Athenodorou:2016oyh,Aguilar:2019uob}, 
\bea
q^2 = p^2= r^2 := s^2\,, \quad q\cdot p = q\cdot r = p\cdot r = -\frac{s^2}{2} \,,  
\label{defsym}
\eea
with the appropriate projector $W^{\alpha\mu\nu}(q,r,p)$ given by Eq.~(2.18) of~\cite{Aguilar:2019uob}.

$L^\sym(s^2)$ receives contributions from both  longitudinal and transverse form factors, 
\be
L^\sym(s^2) =  \underbrace{X_1(s^2) - \frac{s^2}{2}X_3(s^2)}_{L^\sym_{{\scaleto{L}{3pt}}} (s^2)} \, 
+ \underbrace{\frac{s^4}{4}Y_1(s^2) - \frac{s^2}{2}Y_4(s^2)}_{L^\sym_{{\scaleto{T}{3pt}}}(s^2)} \,.
\label{splitL}
\ee
Thus, using the results for $X_i$  obtained in~\cite{Aguilar:2019jsj}, we may express  $L^\sym_{\s L}(s^2)$ as
\be 
L^\sym_{\s L}(s^2) = F(s^2)\left[ J(s^2) \left( H_1(s^2) + \frac{s^2}{2} H_3(s^2) \right) + \frac{s^2}{2} J^{\prime}(s^2) H_2(s^2) \right]\,,
\label{diffeqsym}
\ee
where the $H_i(s^2)$ are the combinations of the $A_i$ and their derivatives given in Eq.~(4.2) of~\cite{Aguilar:2019uob}.

Exactly as \1eq{diffeq_asym}, this last equation
assumes the form of a first order linear differential equation for $J(s^2)$, namely ($x = s^2$)
\be
J^{\prime}(x) + \bsy{P} (x) J(x)  = \bsy{Q}(x)   \,,
\label{ode}
\ee
where 
\be
\bsy{P}(x) = \frac{\bsy{f}_1(x)}{x} \,,\qquad  \bsy{Q}(x) = \frac{\bsy{f}_2(x)}{x} \,,
\label{PandQsym}
\ee
and
\be
\bsy{f}_1(x)  = \frac{2H_1(x)+x H_3(x)}{H_2(x)}\,, \qquad \bsy{f}_2(x) = \frac{2 \left[ L^{\sym}(x) - 
L^\sym_{\s T}(x) \right] }{H_2(x) F (x)} \,.
\label{f1f2}
\ee
Then, the analysis leading to \1eq{condfin_asym} can be repeated, with some minor adjustments. 

From the dynamical equations describing $A_i(s^2)$~\cite{Aguilar:2018csq}, one may show that at the origin
\be
H_1(0) = H_2(0) = Z^\sym_1 \,, \qquad  [ s^2 H_3(s^2) ]_{s^2=0} = 0\,,
\label{Hior_sym}
\ee
where $Z^\sym_1$ is the exact analogue of $Z^\asym_1$. Hence, \1eq{f1f2} implies that $\bsy{f}_1(0) = 2$, whereas in the
asymmetric case we had ${f_1}(0) = 1$; this extra factor of 2 accounts for the double pole encountered below.

Specifically, the solution of \1eq{ode} reads
\be
J(x) = \frac{1}{x^2 \,\bsy{\sigma}(x)}  \left[\mu^4 \bsy{\sigma}(\mu^2)  + \int_{\mu^2}^x \!\!dt \, t \, \bsy{\sigma}(t) \bsy{f}_2(t)\right]\,,
\label{solgen3}
\ee
where
\be
\bsy{\sigma}(x) := \exp\left[\int \! dx\, \bsy{u}(x)/x \right] \,, \qquad \bsy{u}(x) := \bsy{f}_1(x) - 2 \,,
\label{theg}
\ee
and $\bsy{u}(0)=0$.

The main difference between \1eq{solgen3} and \1eq{solgen_asym} is that now the would-be pole in the solution is not
simple but double. Therefore, one needs to consider one more term in the corresponding Taylor expansion, namely  
\be
\mu^4 \bsy{\sigma}(\mu^2)  + \int_{\mu^2}^x \!\!dt \, t \, \bsy{\sigma}(t) \bsy{f}_2(t)
= {\bar a}_0 + {\bar a}_1 x + \frac{1}{2}\, {\bar a}_2 x^2 + {\cal O}(x^3) \,, 
\label{TayExpsy}
\ee
with
\bea
{\bar a}_0 &=& \mu^4 \bsy{\sigma}(\mu^2) + \int_{\mu^2}^0  \!\!\!dt  \, t \, \bsy{\sigma}(t) \bsy{f}_2(t) \,,
\nonumber\\
{\bar a}_1 &=&  [x \bsy{\sigma}(x) \bsy{f}_2(x)]_{x=0}\,,
\nonumber\\
{\bar a}_2 &=& \left[ \bsy{\sigma}(x) \bsy{f}_2(x) + x \bsy{f}^{\prime}_2(x) \bsy{\sigma}(x) +x \bsy{\sigma}^{\prime}(x) \bsy{f}_2(x) \right]_{x=0} \,.
\label{a123}
\eea
Clearly, in order for the solution not to have a pole (double or simple) at the origin, we must have
${\bar a}_0=0$ and ${\bar a}_1=0$.
The first condition amounts to the integral constraint 
\be
\int_0^{\mu^2} \!\!\!dt  \, t \, \bsy{\sigma}(t) \bsy{f}_2(t) = \mu^4 \bsy{\sigma}(\mu^2) \,,
\label{condfin}
\ee
which constitutes the second major result of the present study, 
to be referred to as the ``{\it symmetric sum rule}''.

Given that $\bsy{\sigma}(0) =1$, the condition ${\bar a}_1=0$ 
is satisfied as long  as $\bsy{f}_2(x)$ diverges more mildly than $1/x$; 
evidently, this is comfortably fulfilled by the {\it physical} $\bsy{f}_2(x)$, which diverges logarithmically at the origin. 

If the above two conditions hold, \1eq{solgen3} yields $J(0) = \frac{1}{2}\, {\bar a}_2$; 
its leading contribution is identical (as it should) to that 
obtained  by setting $s^2=0$ directly into \1eq{diffeqsym}.

\section{\label{num_asym} Numerical analysis}

In this section, we focus on the sum rule of~Eq.~\eqref{condfin_asym} and analyze in detail how
it can be used to restrict the form of $V(\ell^2)$, which is 
one of the main ingredients entering in the one-loop dressed approximation of the function ${\cal W}(q^2)$,
given by \2eqs{a1tan_sde}{finalW}.

\subsection{\label{numasym-com} Setting up the stage}

{\it (i)} It is convenient to cast the sum rule~\eqref{condfin_asym} into the equivalent form 
\be
\int^{\mu^2}_0\!\!\! dt\, {\mathcal K}(t)   = 1 \,,
\qquad\quad   {{\mathcal K}}(t):=  \frac{{{\mathcal R}}(t) L^{\asym}(t) }{\mu^2 Z_1^{\asym}F(0)}\,,
\qquad\quad {{\mathcal R}}(t) := {\sigma}(t)/{\sigma}(\mu^2)\,.
\label{norm_asym}
\ee
The quantity ${{\mathcal R}}(t)$ captures the net effect that
different forms of ${\sigma}(t)$ induce on the kernel ${{\mathcal K}}(t)$.

{\it (ii)}  Throughout the analysis, we use \mbox{$\mu= 4.3\,\mbox{GeV}$}, and
\mbox{$\alpha_s(\mu):=g^2(\mu)/4\pi=0.27$}, as determined by the lattice simulation of~\cite{Boucaud:2017obn}.

{\it (iii)} For  $\Delta(q^2)$ and $F(q^2)$,   
 renormalized at the aforementioned $\mu$, we employ the fits
given in Eqs.~(4.1) and  Eq.~(6.1) of~\cite{Aguilar:2018csq}, respectively. Both fits are in excellent agreement with the lattice data of~\cite{Bogolubsky:2007ud}. Note that from Eq.~(6.1) of~\cite{Aguilar:2018csq}, one has \mbox{$F(0)=2.82$}.
Moreover, in accordance with the
discussion of Appendix~\ref{renor}, we set \mbox{$\widetilde{A}_1 = Z^\asym_1 \approx 0.9$}.

\begin{figure}[t]
\includegraphics[scale=0.30]{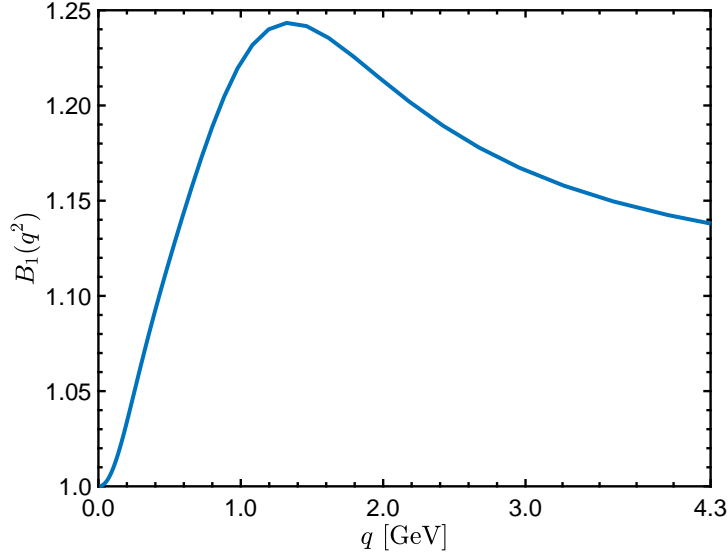}
\caption{ The ghost-gluon vertex form factor, $B_1(q^2)$, given by \1eq{B1}.}
\label{fig:B1_fits}
\end{figure}

{\it (iv)} For $B_1(q^2)$, entering in \1eq{finalW}, 
we use the curve shown  in~\fig{fig:B1_fits}, obtained from
the numerical solution of the SDE governing the ghost-gluon vertex, evaluated 
in the ``soft-ghost'' kinematic limit; see~\cite{Aguilar:2018csq} for details. 
$B_1(q^2)$ can be accurately fitted by   the functional form
 \be 
B_1(q^2)= 1 + \alpha_s \left[ \frac{\omega_1 ( q^2/\kappa_1^2 )^{\gamma_1}}{ 1 + ( q^2/ \kappa_1^2 )^{\gamma_1} } + \frac{ \omega_2 ( q^2/ \kappa_2^2 ) }{ 1 + ( q^2/ \kappa_2^2 )^{\gamma_2} } \right] \,, 
\label{B1}
\ee
where the adjustable parameters acquire the following values:  \mbox{$\gamma_1 = 1.128$},
\mbox{$\gamma_2 = 1.84$}, \mbox{$\kappa_1^2 =  0.101$ GeV$^2$},
\mbox{$\kappa_2^2 = 1.59$ GeV$^2$}, 
\mbox{$\omega_1 = 0.379$}, and \mbox{$\omega_2 = 1.071$}.

\begin{figure}[t]
\includegraphics[scale=0.3]{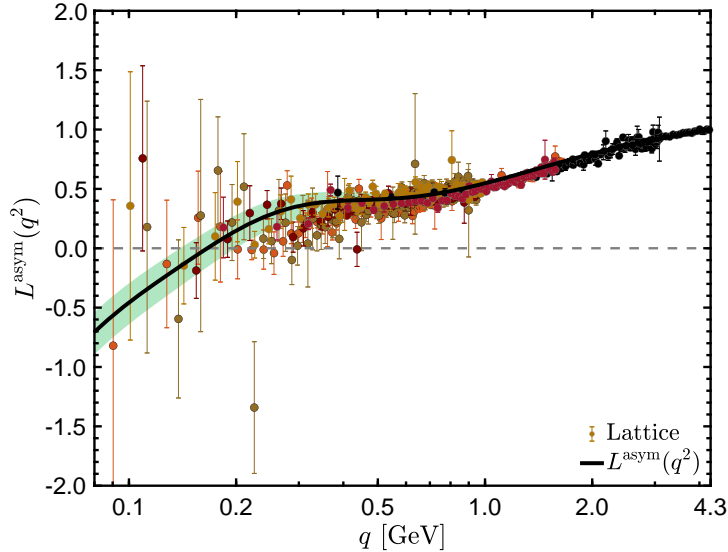}
\caption{Lattice data for $L^\asym(q^2)$ (circles) from~\cite{Athenodorou:2016oyh,Boucaud:2017obn}, and
the fit given by~\1eq{Lasymfit}. The green  shaded band shows a spread in the $L^\asym(q^2)$  of  approximately  $0.37$ at zero momentum,  and reduces to $0.20$  at \mbox{$300$ MeV}.}
\label{fig:Lasym_vars}
\end{figure}

{\it (v)} For  $L^\asym(q^2)$ we employ a rather good
fit to the lattice  data of~\cite{Athenodorou:2016oyh,Boucaud:2017obn} whose $\chi^2/{\rm d.o.f}=0.024$. The
curve is shown in Fig.~\ref{fig:Lasym_vars}, and its  functional form is given by  
\be 
L^\asym(q^2) = F(q^2)T(q^2) + \nu_1 \left( \frac{1}{ 1 +  ( q^2 /\nu_2 )^2 } - \frac{1}{ 1 +  ( \mu^2 /\nu_2 )^2 } \right) \,,
\label{Lasymfit}
\ee
with   
\be 
T(q^2) = 1 + \frac{ 3 \lambda_2 }{ 4 \pi }\left( 1 + \frac{ \tau_1 }{ q^2 + \tau_2 } \right) \left[ 2 \ln\left( \frac{q^2 + \eta^2(q^2)}{\mu^2} \right) + \frac{1}{6}\ln\left( \frac{q^2}{\mu^2}\right) \right] \,, 
\label{kinetic_fit}
\ee
and  
\be 
\eta^2(q^2) = \frac{\xi}{q^2 + \tau_0 } \,, 
\label{eta}
\ee
where the fitting parameters are \mbox{$\lambda_2=0.276$},   \mbox{$\tau_0=0.41$ GeV$^2$}, \mbox{$\tau_1 = 4.05$ GeV$^2$}, \mbox{ $\tau_2 = 0.16$ GeV$^2$ }, \mbox{$\nu_1 = 0.52$}, \mbox{$\nu_2 = 0.012$ GeV$^2$}, and  $\xi =10.2\,\mbox{GeV}^4$.

Note that the above fit incorporates, by construction, the renormalization condition 
\mbox{$L^\asym(\mu^2)=1$}, corresponding to the
``asymmetric'' MOM scheme employed. 
In addition, the zero crossing of $L^\asym(q^2)$ is  located at   \mbox{$167$ MeV}. 

Observe that the above functional form captures the expected infrared asymptotic
behavior of Eq.~\eqref{condor_asym}. In particular,  by expanding  Eq.~\eqref{Lasymfit} around \mbox{$q^2\to 0 $},
one obtains
\be
\lim\limits_{q^2\to 0} L^\asym(q^2)  =    a \ln(q^2/\mu^2) + b \,,
\label{asym}
\ee
where $a$ and $b$ are constants, comprised by combinations of
the fitting parameters entering in the Eq.~\eqref{Lasymfit};  
the validity of~\eqref{asym} is restricted to the range \mbox{($0-80$) MeV}.

In Fig.~\ref{fig:Lasym_vars},  the green shaded area is obtained by varying $\xi$ by $\pm 2\%$,
keeping all other fitting parameters fixed; equivalently, at the level of~\eqref{asym},  this amounts to
changing $b$ by $\pm 5\%$, while $a$ remains unchanged.
The spread induced to $L^\asym(q^2)$  by this variation is more pronounced in the infrared [ $0.37$ at \mbox{$100$ MeV}],
being gradually reduced as one moves towards higher momenta [\eg $0.20$  at \mbox{$300$ MeV}].

{\it (vi)}  To every set of ingredients entering into the sum rule 
we will assign the corresponding percentage error, denoted by $\varepsilon$, measuring the deviation of the result from unity, \ie
\be
\varepsilon = \left[\int^{\mu^2}_0 \!\!\!\!\! dt \,{{\mathcal K}}(t) -1 \right]\times 100\%\,.
\label{error}
\ee

{\it (vii)} The function ${\sigma}(x)$ is defined in \1eq{theg_asym} in terms of an indefinite integral; however,
 if the integration cannot be carried out analytically, 
 one needs to convert it to a definite integral, to be evaluated numerically. 
This, in turn, introduces an arbitrary integration constant,
which amounts to a rescaling of the answer; nonetheless, it is clear
from \2eqs{solgen3}{condfin} that such a rescaling is irrelevant, given that ${\sigma}(x)$  enters homogeneously in the sum rule.
In what follows we use the form
\be
{\sigma}(x) = \exp\left[\int_{0}^{x} \!\! dt\, {u}(t)/t \right];
\label{sigmadef}
\ee
choosing this particular lower limit of integration fixes the overall scale such that ${\sigma}(0) =1$.

\subsection{\label{numasym-rest} Pinning down $V(\ell^2)$}


We next present a concrete example of how the sum rule of \1eq{norm_asym},
accompanied by a set of physically motivated assumptions, may restrict severely some of 
the ingredients comprising its kernel. Specifically, 
we discuss in detail the  impact 
that the variations  of the form factor $V(\ell^2)$ have on ${\cal W}(q^2)$, and, eventually, 
through the form of ${\sigma}(q^2)$, on the  sum rule itself.
This particular choice is prompted by the numerical exploration of the expressions for
${\cal W}_{d_1}(q^2)$ and ${\cal W}_{d_2}(q^2)$, given in \1eq{finalW}, whose upshot is  that  
${\cal W}_{d_2}(q^2)$ furnishes the dominant contribution, and that its value is considerably more sensitive
to variations of $V(\ell^2)$ rather than of $B_1(q^2)$ [$F(q^2)$ and $\Delta(q^2)$ are held fixed]. 
It is therefore reasonable to establish whether the sum rule is able to place nontrivial bounds on the form
and main features exhibited by $V(\ell^2)$.

Let us emphasize at this point that $V(\ell^2)$ emerged at the final step of a series of approximations, described
in Appendix~\ref{a1tan_deriv}, whose purpose was to simplify the treatment of the equations defining ${\cal W}(q^2)$; 
it should be therefore interpreted as an ``effective'' form factor, capturing the collective
action of the various $X_i$  and $Y_i$ comprising the three-gluon vertex, with their multitude of sizes and kinematic dependence. 
In this sense, there is no a {\it priori} guarantee that 
$V(\ell^2)$ will inherit from them their characteristic suppression in the intermediate region of momenta.
Nonetheless, as we will see in what follows, the sum rule clearly favors a ``suppressed'' $V(\ell^2)$,
imposing, at the same time, strict limits on the amount of its suppression.

\newcolumntype{g}{>{\columncolor[HTML]{d7dbdd}}c}
\begin{table}[t]
\begin{center}
\begin{tabular}{|g|c|c|c|c|c|}
\toprule
$V(\ell^2) $ & $\quad V_{1}(\ell^2)\quad$ & $\quad V_{2}(\ell^2)\quad$ & $ \quad V_{3}=1\quad$ & $ \quad V_{\!\star}(\ell^2)\quad$ & $ \quad V_{4}(\ell^2)\quad$     \\
\hline
$\xi$ [GeV$^{4}$] & $266.4$ & $169.5$ & $-$ & $99.4$& $24.2$  \\
 \hline
 $\tau_0$ [GeV$^{2}$]& $10.6$ & $7.3$ & $-$ & $4.8$& $1.2$  \\
\hline
$\varepsilon (\%)$ & $24.5$ & $13.0$ & $9.5$ & $0$& $-19.5$  \\
\hline
\end{tabular}
\end{center}
\caption{ \label{xi_parameter}  The values of the parameters $\xi$ and $\tau_0$ used  
in~\eqref{X1fit} in order to obtain the   
$V_i(\ell^2)$ shown in Fig.~\ref{fig:X1_fits}.  The tree-level case, $V_{\s 3} = 1$,
is recovered by simply setting \mbox{$\lambda_s=0$}. In the last row we quote
the percentage error $\varepsilon$, given by Eq.~\eqref{error}, when  the sum rule~\eqref{norm_asym}  is computed using the ${\sigma}_i(q^2)$ obtained with the  
variations  of the form factor $V_i(\ell^2)$ shown in Fig.~\ref{fig:X1_fits}.}
\end{table}

\begin{figure}[t]
\includegraphics[scale=0.3]{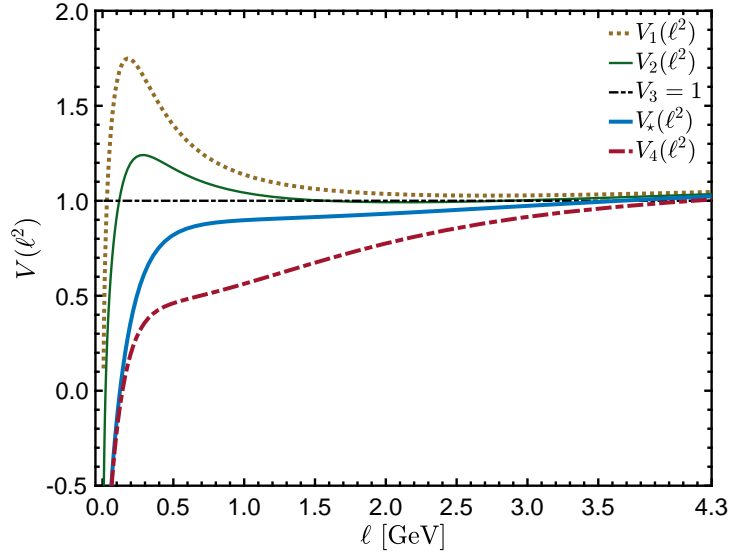}
\caption{ The five representative cases for $V_i(\ell^2)$. All curves are obtained using the the {\it Ansatz} given by  Eq.~\eqref{X1fit}, together with the set of parameters quoted in Table~\ref{xi_parameter}.}
\label{fig:X1_fits}
\end{figure}

Next, we introduce a concrete functional form for $V(\ell^2)$,
whose variations will generate distinct versions of this quantity. Specifically, we employ the {\it Ansatz}
\be 
V(\ell^2) = 1 + \frac{\lambda_s}{ 32 \pi }\left( 1 + \frac{\tau_1}{ \ell^2 + \tau_2 } \right)\left[ 33 \ln\left( \frac{\ell^2 + \eta^2(\ell^2) }{\mu^2} \right) + \ln\left( \frac{\ell^2}{\mu^2} \right) \right] \,,
\label{X1fit}
\ee
with $\eta^2(\ell^2)$ given by  Eq.~\eqref{eta}. The parameters that will be varied throughout this analysis
are $\xi$  and $\tau_0$; all others are kept fixed at the values
\mbox{$\lambda_s = 0.22$},  \mbox{$\tau_1 = 11.6$ GeV$^2$}, and \mbox{$\tau_2 = 0.0856$ GeV$^2$}.
In Table~\ref{xi_parameter} we quote the sets $\{\xi, \tau_0\}$ that generate  
the curves shown in Fig.~\ref{fig:X1_fits}. Note that, depending 
on the choice of the fitting parameters, one may achieve {\it suppression or enhancement} in
the intermediate region of momenta. Moreover, as commented below,
the infrared and ultraviolet limits of this {\it Ansatz} are motivated by theoretical considerations.

The starting point of the study consists in establishing the response of the sum rule in two relatively ``extreme'' 
situations: $V(\ell^2)=V_{1}(\ell^2)$, showing a substantial enhancement, and $V(\ell^2) = V_{4}(\ell^2)$, displaying
a considerable amount of suppression. The substitution of these two versions of $V(\ell^2)$ into 
the formulas furnishing ${\cal W}(q^2)$ [\2eqs{a1tan_sde}{finalW}], and the subsequent evaluation of \1eq{error}, 
reveals that the resulting errors differ in sign: \mbox{$\varepsilon_{\s 1}>0$}, but \mbox{$\varepsilon_{\s 4}<0$} [see Table~\ref{xi_parameter}]. 

Since $\varepsilon$ changes sign when switching from  \mbox{$V(\ell^2)=V_{1}(\ell^2)$} to \mbox{$V(\ell^2)=V_{4}(\ell^2)$}, 
it is reasonable to expect that 
there will be an ``intermediate'' curve, to be
denoted by $V_{\!\star}(\ell^2)$,  for  which the sum rule will be fulfilled exactly (\mbox{$\varepsilon=0$}).
To determine it, we employ \1eq{X1fit} in order to produce
a sequence of variations for $V(\ell^2)$, determining each time the corresponding $\varepsilon$; note that  
the modifications to $V(\ell^2)$ are implemented  
in the transition  region between nonperturbative and perturbative regimes,
\ie approximately  from  \mbox{$300$ MeV}  to  \mbox{$3$ GeV}, as shown in the Fig.~\ref{fig:X1_fits}.
All variations considered merge together in the ultraviolet and infrared regimes; 
specifically, in the ultraviolet they recover the  one-loop results 
for \mbox{$X_1(\ell^2)= X_4(\ell^2)= X_7(\ell^2)$}~\cite{Ball:1980ax,Davydychev:1996pb}, while in the infrared
they diverge at a common logarithmic rate~\cite{Aguilar:2019jsj}.

In Table~\ref{xi_parameter} we show the values for $\varepsilon$ 
when the sum rule is evaluated using the ${\sigma}(q^2)$ 
obtained  with the various $V_i(\ell^2)$ shown in Fig.~\ref{fig:X1_fits}.
Evidently, the sum rule favors clearly a \mbox{$V(\ell^2)=V_{\!\star}(\ell^2)$} that is {\it suppressed} with respect to the tree-level value captured by $V_{3}$,  but certainly less so than the initial $V_{4}(\ell^2)$.

\begin{figure}[t]
\begin{minipage}[b]{0.45\linewidth}
\hspace{-1.5cm}
\centering
\includegraphics[scale=0.27]{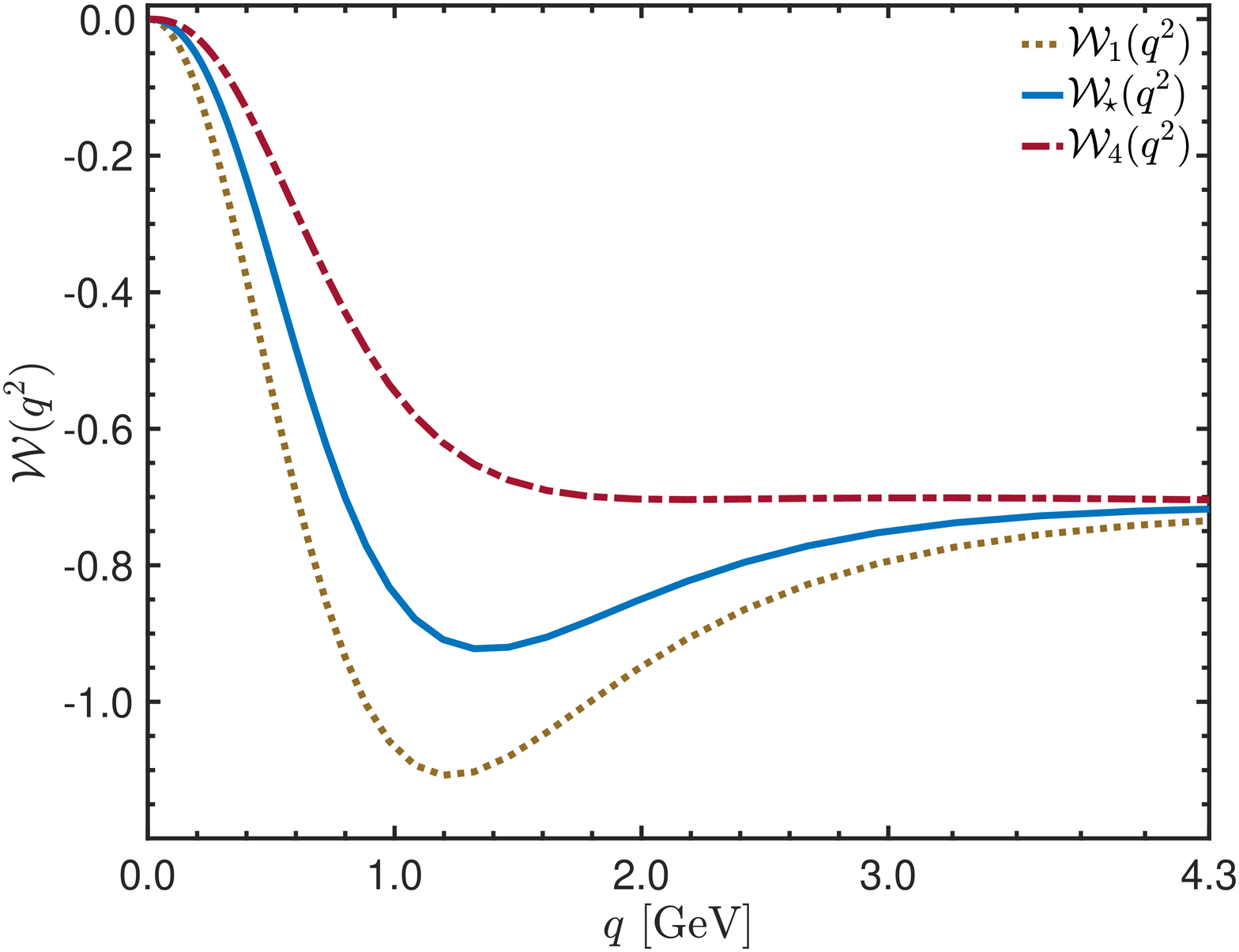}
\end{minipage}
\hspace{0.15cm}
\begin{minipage}[b]{0.45\linewidth}
\includegraphics[scale=0.27]{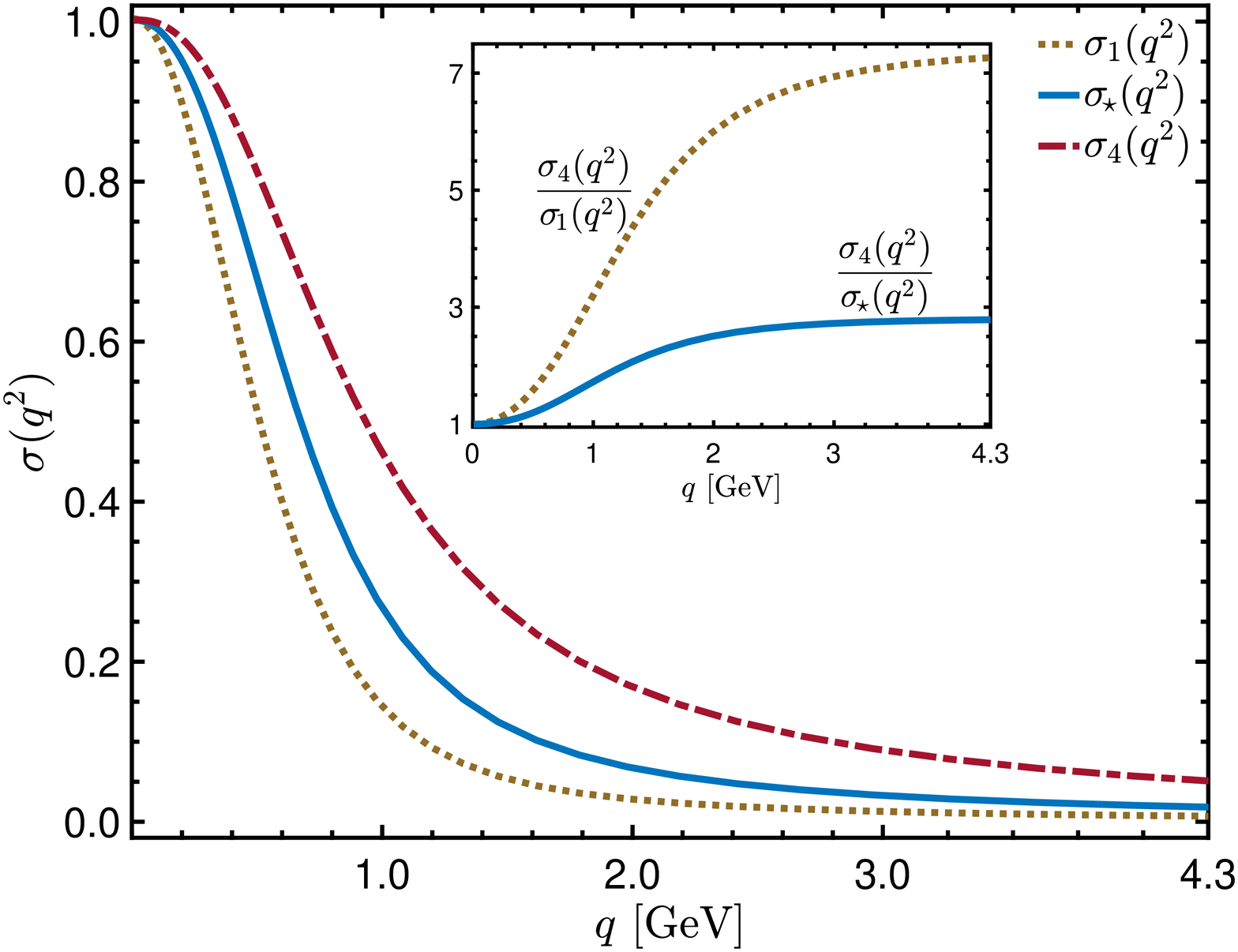}
\end{minipage} 
\begin{minipage}[b]{0.45\linewidth}
\hspace{-1.5cm}
\centering
\includegraphics[scale=0.27]{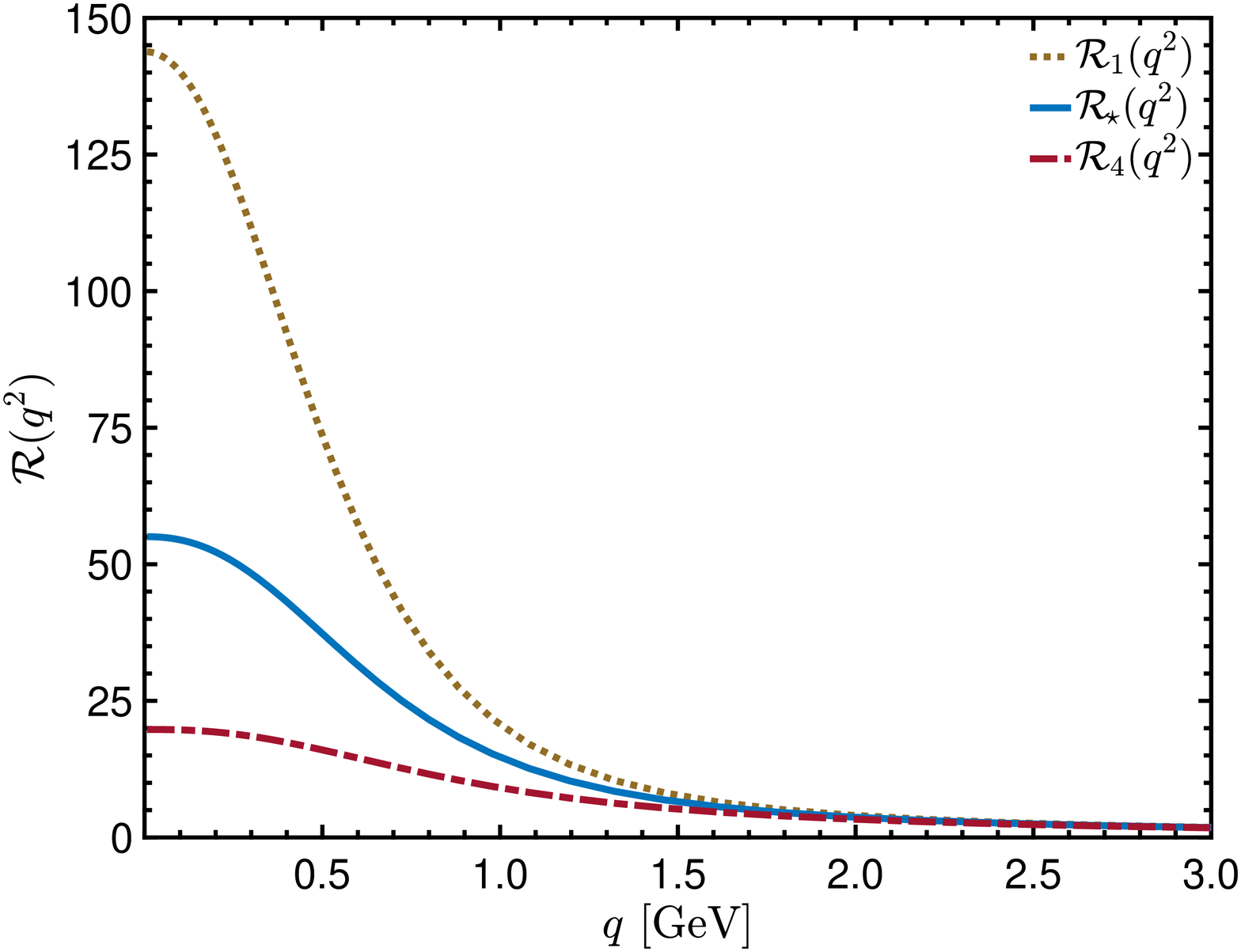}
\end{minipage}
\hspace{0.15cm}
\begin{minipage}[b]{0.45\linewidth}
\includegraphics[scale=0.27]{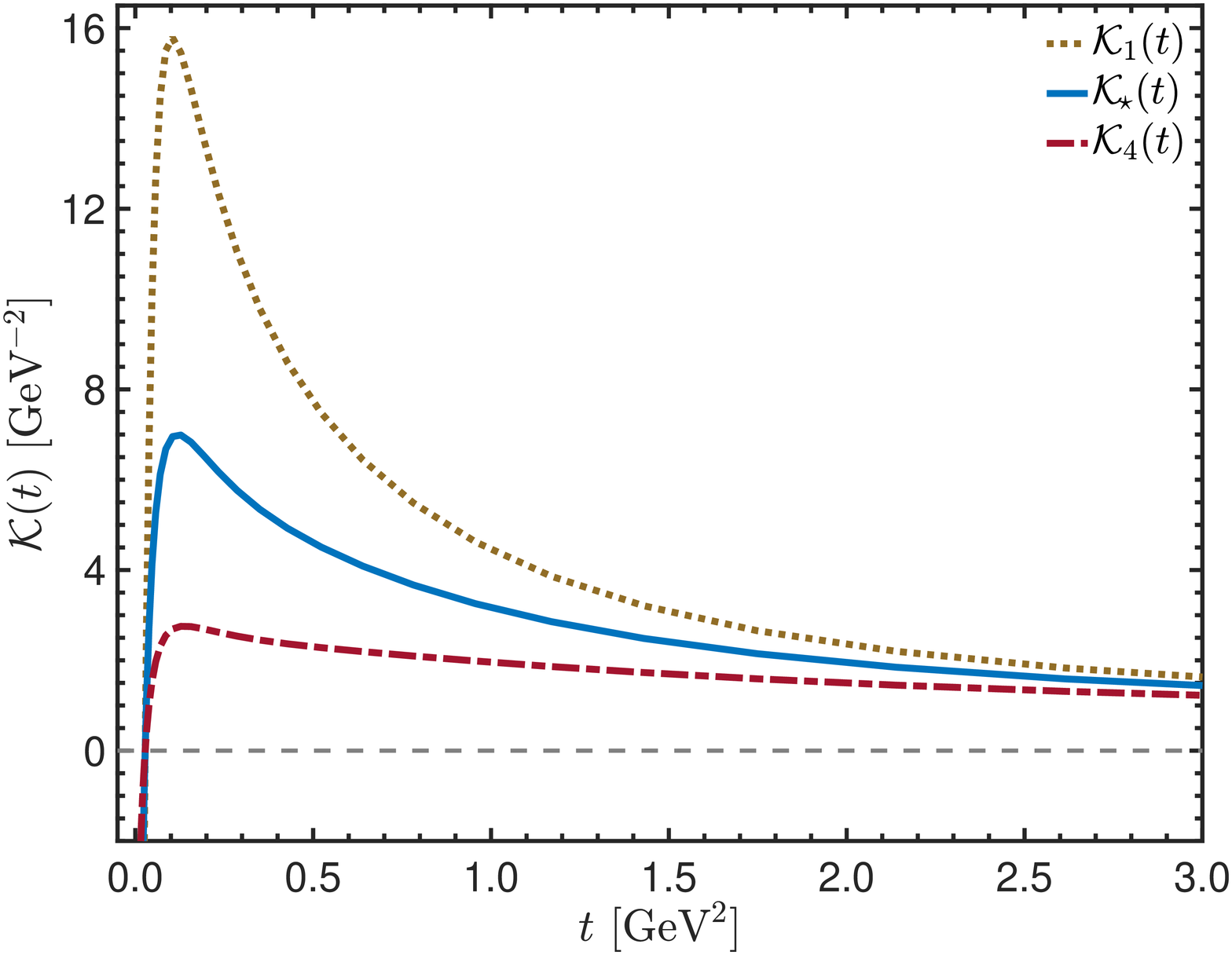}
\end{minipage} 
\caption{Top left panel:   The functions ${\cal W}_i(q^2)$  obtained using  $V_{1}(\ell^2)$ (brown dotted), 
$V_{\!\star}(\ell^2)$ (blue continuous), and $V_{4}(\ell^2)$ (red dashed-dotted), shown in Fig.~\ref{fig:X1_fits}.
Top right panel:   The corresponding 
  ${\sigma}_{1}(q^2)$, ${\sigma}_{\!\star}(q^2)$, and ${\sigma}_{4}(q^2)$.  Bottom left panel: The ratios ${\mathcal R}_{1}(q^2)$,  ${\mathcal R}_{\!\star}(q^2)$,  and  ${\mathcal R}_{4}(q^2)$,   defined in  Eq.~\eqref{norm_asym}. 
Bottom right panel: The integrands ${{\mathcal K}}_{1}(t)$,  ${{\mathcal K}}_{\!\star}(t)$, and 
${{\mathcal K}}_{4}(t)$,  given by Eq.~\eqref{norm_asym}. }
\label{fig:w_g_asym}
\end{figure}

It is rather instructive to analyze in some detail how the above result emerges.
To that end, denote by \mbox{[${\cal W}_{1}(q^2)$, ${\sigma}_{1}(q^2)$]}, \mbox{[${\cal W}_{\!\star}(q^2)$, ${\sigma}_{\!\star}(q^2)$]}, and \mbox{[${\cal W}_{4}(q^2)$, ${\sigma}_{4}(q^2)$]}, the corresponding quantities
obtained when \mbox{$V(\ell^2)=V_{1}(\ell^2)$},  \mbox{$V(\ell^2)=V_{\!\star}(\ell^2)$}, and \mbox{$V(\ell^2)=V_{4}(\ell^2)$}, respectively. 
As can be seen in Fig.~\ref{fig:w_g_asym}, the suppression induced to $V(\ell^2)$ through the transition 
\mbox{$V_{1}(\ell^2) \to V_{4}(\ell^2)$}
leads to an enhancement of 
${\cal W}(q^2)$  in the region of \mbox{$800$ MeV} - \mbox{$2$ GeV},
where clearly  \mbox{${\cal W}_{4} (q^2)> {\cal W}_{1}(q^2)$}. This difference is transmitted to the
${\sigma}_{1}(q^2)$ and ${\sigma}_{4}(q^2)$, which, as depicted in  Fig.~\ref{fig:w_g_asym},
satisfy the relation \mbox{${\sigma}_{4}(q^2) > {\sigma}_{1}(q^2)$}  in the entire range of momenta. 
Notice that, 
even though ${\cal W}_{1}(q^2)$ and ${\cal W}_{4}(q^2)$ merge into each other past $q=4$ GeV,
${\sigma}_{1}(q^2)$ and ${\sigma}_{4}(q^2)$ remain clearly separated in the same region of momenta; in fact, as can be seen in the inset, they reach their maximum difference precisely at the end point of the momentum interval. 
This, in turn, indicates that the ultraviolet behavior of ${\sigma}(q^2)$ is particularly sensitive
to the low energy structure of the corresponding ${\cal W}(q^2)$. 
Quite interestingly, as can be seen in Fig.~\ref{fig:w_g_asym}, when the corresponding ratios   
${{\mathcal R}}_{1}(q^2)$ and ${{\mathcal R}}_{4}(q^2)$ are formed, 
the initial hierarchy ${\sigma}_{4}(q^2) > {\sigma}_{1}(q^2)$ is inverted:
now ${{\mathcal R}}_{4}(q^2)< {{\mathcal R}}_{1}(q^2)$.
As a result, the suppressed $V_{4}(\ell^2)$ gives rise to the
kernel ${{\mathcal K}}_{4}(t)$ that is itself 
suppressed, with respect to ${{\mathcal K}}_{1}(t)$, in the momentum interval 
between the zero crossing (at approximately $0.03$ GeV$^2$) and $\mu^2$,
which practically accounts for the entire value of the sum rule integral.  
Indeed, the support between the origin and the zero crossing
is completely negligible; for example, 
in the case of ${{\mathcal K}}_{\!\star}(t)$, it contributes to the
full answer a mere $-0.017$.
Consequently, the logarithmic divergence encoded into $L^{\asym}(q^2)$ is practically undetected by the sum rule. 
%
%

\begin{figure}[t]
\includegraphics[scale=0.30]{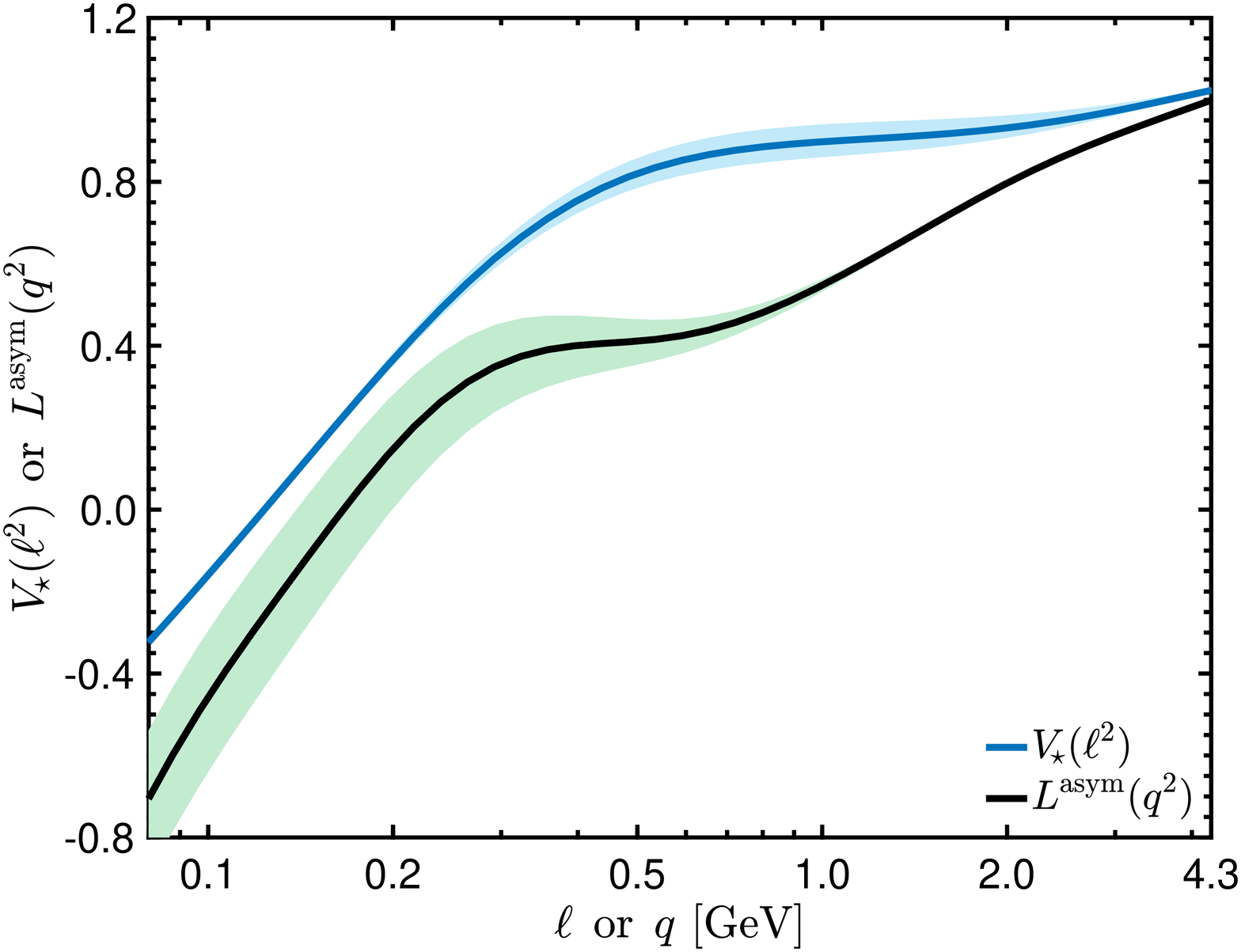}
\caption{ The propagation of the uncertainty in the $L^\asym(q^2)$ to the $V_{\!\star}(\ell^2)$, 
through the action of the  sum rule.}
\label{fig:varyLi}
\end{figure}

Up until this point, the entire analysis has been carried out using a {\it fixed} $L^{\asym}(q^2)$,
namely the one given by \1eq{Lasymfit},
with the values of the parameters quoted below \1eq{eta}.
It is clearly important to examine the stability of the result obtained 
for $V_{\!\star}(\ell^2)$ when variations in the form of $L^{\asym}(q^2)$
are introduced. To that end, we repeat the previous procedure, using for $L^{\asym}(q^2)$ the two limiting curves that demarcate the green shaded area 
in Fig.~\ref{fig:Lasym_vars}; the corresponding $V_{\!\star}(\ell^2)$ turn out to be
particularly close to the original one, forming the narrow blue band shown in Fig.~\ref{fig:varyLi}.

In Fig.~\ref{fig:varyLi} one may appreciate how the sum rule maps 
the original uncertainty in $L^{\asym}(q^2)$ (green band) into 
a corresponding uncertainty in $V_{\!\star}(\ell^2)$ (blue band). In particular, note that, through the action of the
function ${\sigma}(q^2)$,  
the maximum separation is shifted from the deep infrared to the
physically more interesting range of \mbox{$300\,\mbox{MeV} - 1.3\,\mbox{GeV}$}, where the three
$V_{\!\star}(\ell^2)$ reach their maximum mutual discrepancy of about $4\%$.

\section{\label{sec:conc} Discussion and Conclusions }

We have presented a couple of new sum rules,
which originate from the STIs that connect the two- and three-point
sectors of quenched QCD, in the Landau gauge.
The key observations that are crucial for their derivation may be summarized as follows.
In the context of two special kinematic configurations that involve a single momentum scale, 
the nonperturbative BC solutions may be interpreted as exactly solvable ordinary differential equations 
for the function  $J(q^2)$, known as the ``kinetic'' term of the gluon propagator.
The general solutions of these differential equations predict
the presence of poles (simple or double) at the origin; however, as dictated by its own SDE,
$J(q^2)$ must diverge at the origin only logarithmically. Nonetheless, as an appropriate expansion reveals, 
these two descriptions may be eventually reconciled, provided
that certain integral conditions (sum rules) are exactly satisfied.
These sum rules are referred to as ``asymmetric'' or ``symmetric'', depending on the kinematic configuration that
has served as the starting point for their derivation.

The only element from the two-point sector that enters in these sum rules 
is the ghost dressing function; its behavior
is particularly well-established, thanks to large-volume lattice simulation as well as a variety of continuous approaches.
All remaining ingredients are related to the three-point sector, 
whose quantitative exploration, despite the considerable advances mentioned in the Introduction, remains a major technical challenge for the QCD practitioners.
The sum rules may serve as a complementary tool in this ongoing quest, 
furnishing useful constraints for the various components entering in them. 

For the purposes of the present work, we have  
opted for a mixed approach, where certain of the quantities, such as $L^\asym(q^2)$, 
were obtained from the lattice, while the components related to the ghost-gluon kernel from the corresponding
SDEs. Alternatively, one may resort to an entirely SDE-based analysis, along the lines of~\cite{Schleifenbaum:2004id,Huber:2012kd,Aguilar:2013xqa,Huber:2012zj,Blum:2014gna,Eichmann:2014xya,Williams:2015cvx},
or to an exclusive functional renormalization group treatment, in the spirit of~\cite{Pawlowski:2005xe,Cyrol:2016tym,Cyrol:2017ewj}, such that all relevant quantities are computed within a self-contained framework,
and their quality is subsequently assessed by means of the corresponding sum rule.
In that vein, one may also envisage a purely lattice-driven approach, 
especially in the context of the asymmetric sum rule; such an effort would entail the
simulation of the ghost-gluon kernel, and the extraction of the function ${\cal W}(q^2)$.

In the numerical study of Sec.~\ref{num_asym} we have used the asymmetric sum rule
in order to constrain the effective form factor $V(\ell^2)$, entering in the
SDE that determines ${\cal W}(q^2)$. In that sense, the sequence 
$V(\ell^2) \to {\cal W}(q^2) \to \varepsilon$ was considered, and the ``perfect''  
$V_{\!\star}(\ell^2)$ ($\varepsilon=0$) has been determined, which
displays a distinct suppression (with respect to unity) in the intermediate range of momenta.
It is important to emphasize that {\it no initial bias} regarding the suppression of $V(\ell^2)$ was built in;
in fact, cases with notable enhancement have been considered ($V_1$, $V_2$) , which, however, were clearly disfavored by the sum rule, as was the case with excessive suppression ($V_4$). 
Note finally, that, as mentioned at the end of Sec.~\ref{num_asym}, the sum rules are completely insensitive to the presence of the
zero crossing, and the subsequent logarithmic divergence, displayed by all the $L^{\asym}(q^2)$ considered. 

Within the confines of the present approach,
the lack of information on $L^\sym_{\s T}(s^2)$
reduces the usefulness of the symmetric sum rule.
The situation could be amended if $L^\sym_{\s T}(s^2)$ were to be
estimated from the SDE of the three-gluon vertex; in such a case, due to the presence of the extra factor of $t$
in its integrand, the sum rule would be probing the quantities involved predominantly
in the intermediate and  high momentum regimes.

Finally, it would be particularly interesting to return to the starting point of this
investigation, and determine the behavior of the quantity $J(q^2)$ from the 
solution of the differential equation given in \1eq{Jagain}.
From the conceptual point of view, such a construction may be interpreted as an ``inverse'' gauge technique,
in the sense that, instead of determining a vertex from a propagator, as is usually the case, 
now the kinetic term of the gluon propagator would be determined from ingredients of the three-point sector of the theory. 
We expect to present the results of this analysis in the near future.

\appendix

\section{\label{renor} Changing renormalization schemes}

In this Appendix  
we address certain technical issues
related to the renormalization schemes employed in the
computation of the various quantities entering in the sum rule.

{\it  (i)} The  sum  rule of  \1eq{condfin_asym} has  been
derived using {\it explicitly} the condition $J(\mu^2) = 1$, with no
further assumptions regarding the 
renormalization prescription imposed on  the  remaining
quantities. Therefore,  it is valid in the context
of any self-consistent renormalization scheme that accommodates this particular condition;
of course, in practice, one is limited to the few choices 
used in the literature, such as the various MOM variants, as well as the Taylor scheme.

{\it  (ii)}
Since $J(q^2)$, $F(q^2)$, $H_{\nu\mu}$, and $\Gnp_{\alpha\mu\nu}$ are related by the STI of \1eq{stig}, the number of 
renormalization conditions that may be chosen freely is reduced down to three.
Usually, in addition to $J(\mu^2) = 1$, 
one imposes the condition $F(\mu^2)=1$,  
because lattice simulations of two point functions opt for this natural condition,
and the various functional treatments comply, in order to facilitate the comparison
of the results.
Moreover, the lattice data for the special {\it projection} of $\Gnp_{\alpha\mu\nu}$ that we consider
have been renormalized such that $L^\asym(\mu^2) = 1$; we will refer to this particular scheme as ``asymmetric'' MOM.
Thus, at this point, within this particular scheme, the value of the $H_{\nu\mu}$ at the renormalization point $\mu$
is completely fixed by the aforementioned STI.

{\it  (iii)} The practical upshot of these considerations is that for the numerical evaluation of the sum rules one may {\it not} 
use the $A_i$ obtained in~\cite{Aguilar:2018csq} {\it together} with the lattice results 
for $L^\asym(q^2)$. This is so because in~\cite{Aguilar:2018csq}   
the renormalization was carried out in the Taylor scheme~\cite{Boucaud:2008gn}, which requires that $H_{\nu\mu}(q,p,r)$ collapses to    
its tree level value, $g_{\nu\mu}$ in the soft-ghost kinematics ($p=0$); however, this value 
does {\it not} coincide with the one obtained from  
the STI when the asymmetric MOM scheme is employed.
To remedy this inconsistency, the $A_i$ of~\cite{Aguilar:2018csq} must undergo an appropriate rescaling,
which  will render them compatible with all other inputs.

{\it  (iv)}
As is well-known, the choice of renormalization scheme affects the finite part
of the various cutoff-dependent renormalization constants, $Z_i$, 
and the transition of the Green's functions from one scheme to the next may be conveniently described as the action 
of additional finite renormalizations constants.
In that sense, the $H_{\nu\mu}$ calculated in the {\it Landau gauge} is special, because its quantum corrections are known to be {\it finite},
and do not require an infinite renormalization. 
Therefore, the corresponding renormalization constant, $Z_1$, is finite,
and its numerical value may be directly used to describe the transition between the various schemes.
Note in particular, that in the Taylor scheme, which will serve as our point of departure, $Z_1$ acquires the
special value $Z_1=1$~\cite{Taylor:1971ff}. Consequently, the transitions to other schemes will manifest themselves as deviations of $Z_1$ from unity. 
Clearly, the conversion of the $A_i$ into the asymmetric MOM scheme requires the use of such a constant, 
to be denoted by $Z^\asym_1$.

{\it  (iv)} We next determine approximately the value of $Z^\asym_1$.  
Reserving the notation $A_1(q^2)$ and ${\cal W}(q^2)$ for the quantities defined in the asymmetric scheme, and denoting
by $A^{\s{\rm T}}_1(q^2)$ and ${\cal W}^{\s{\rm T}}(q^2)$ their counterparts in the Taylor scheme, we have
\be 
\widetilde{A}_1 = Z^\asym_1 \widetilde{A}_1^{\s{\rm T}} = Z^\asym_1\,, \qquad {\cal W}(q^2) = Z^\asym_1 {\cal W}^{\s{\rm T}}(q^2) \,,
\label{z2_def}
\ee
where we used the special result of the Taylor scheme, $\widetilde{A}_1^{\s{\rm T}} = 1$.

Then, we set $q^2 = \mu^2$ in \1eq{diffeq_asym}, and evaluate it in the asymmetric and Taylor schemes; denoting the corresponding results
by $L^\asym(\mu^2)$ and $\bar{L}^\asym(\mu^2)$, respectively, we obtain 
\bea
L^\asym(\mu^2) &=& Z^\asym_1 F(0)\left[ 1 + {\cal W}^{\s{\rm T}}( \mu^2 ) + \mu^2 J'(\mu^2) \right]\,,
\nonumber\\
\bar{L}^\asym(\mu^2) &=& F(0) \left[ 1 + {\cal W}^{\s{\rm T}}( \mu^2 ) + \mu^2 J'(\mu^2) \right] \,.
\label{diffeq_asym_Tay}
\eea

Since, by definition, $L^\asym(\mu^2) = 1$, taking the ratio of the two sides of \1eq{diffeq_asym_Tay} yields 
\be 
Z^\asym_1 = \frac{1}{\bar{L}^\asym(\mu^2)} \,.
\label{z2_complete}
\ee

Next,  we assume that the renormalization point \mbox{$\mu = 4.3$ GeV} used in the lattice
simulations is sufficiently large for perturbation theory to be a reasonable approximation for 
$Z^\asym_1$.
Then, the one-loop results for the $X_i(q,-q,0)$ of~\cite{Davydychev:1996pb,Aguilar:2019jsj}, which were computed in the Taylor scheme, can be used to approximate $\bar{L}^\asym(\mu^2)$ by 
\be 
\bar{L}^\asym(\mu^2) \approx  1 + \frac{37 C_\mathrm{A} \alpha_s }{96\pi} \,,
\ee
where $C_\mathrm{A}$ is the Casimir eigenvalue in the adjoint representation [$N$ for $SU(N)$]. Substituting the above expression in \1eq{z2_complete}, yields ($\alpha_s = 0.27$)
\be 
Z^\asym_1 \approx 1 - \frac{37 C_\mathrm{A} \alpha_s }{96\pi} \approx 0.9 \,.
\label{zasymnum}
\ee

{\it  (v)} 
An exactly analogous discussion holds for the symmetric configuration. 
The condition $L^\sym(\mu^2) = 1$, employed for the lattice data in~\cite{Athenodorou:2016oyh,Aguilar:2019uob},
defines a scheme that is {\it distinct} to the asymmetric MOM, mentioned above; 
by analogy, it is  referred to as the ``symmetric'' MOM scheme.  
The conversion of the $A_i$ from the Taylor to this latter scheme proceeds by means of a 
finite renormalization constant, to be denoted by $Z^\sym_1$, whose numerical value is not required 
in the present work.

\section{\label{a1tan_deriv} Computing the function ${\cal W}(q^2)$}

In this Appendix we provide details related with the determination of ${\cal W}(q^2)$.

\begin{figure}[!h]
\includegraphics[scale=0.45]{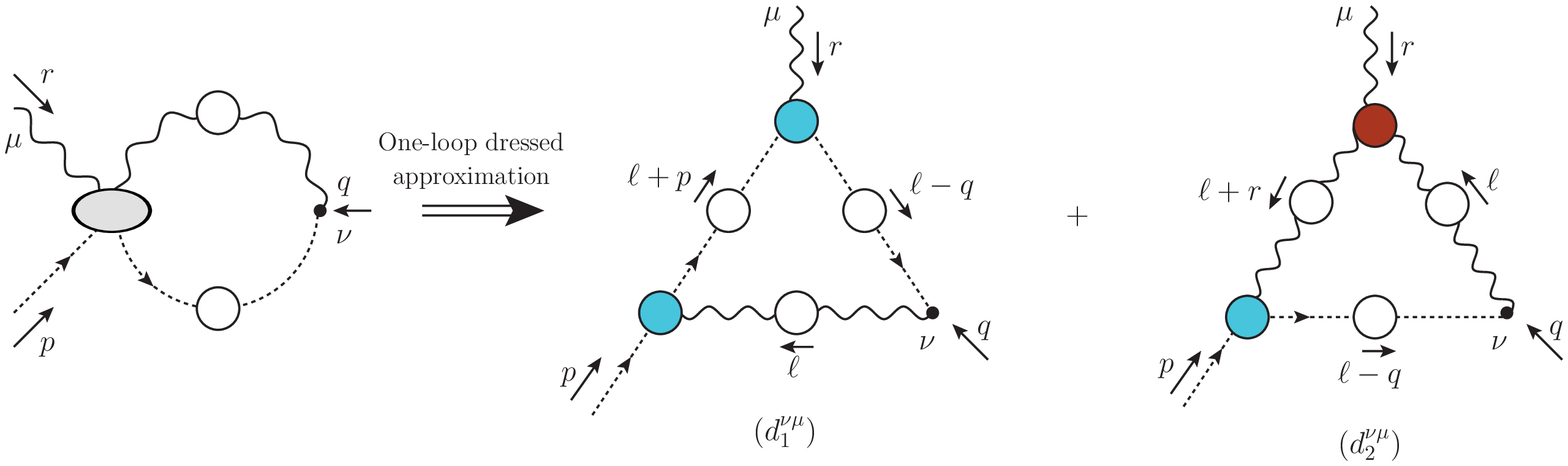}
\caption{One-loop dressed approximation of the SDE governing the ghost-gluon scattering kernel.}
 \label{fig:H_truncated}
\end{figure}

The one-loop dressed  
approximation of $H_{\nu\mu}$ is obtained from the two diagrams of Fig.~\ref{fig:H_truncated} [Eq.~(3.1) of~\cite{Aguilar:2018csq}], 
which are proportional to the ghost momentum $p$, as can be easily established by
contracting the structures $(\ell + p)_{\alpha}$ and $(\ell -q)_{\alpha}$ by the
Landau gauge propagators  $\Delta_{\alpha\beta}(\ell)$ and $\Delta_{\alpha\beta}(\ell+r)$, respectively; their remainder defines
the quantity $K_{\nu\mu\rho}(q,p,r)$, appearing in the Eq.~\eqref{H_tay}, where we set directly $p = 0$. Then, 
from \1eq{HKtens} it is straightforward to deduce that ${\cal W}(q^2)$ can be extracted from
$K_{\nu\mu\rho}(q,0,-q)$ through the projection 

\be 
{\cal W}(q^2) = - \frac{1}{ 3 } \, q^\rho \, {\rm P}^{\mu\nu}(q) \, K_{\nu\mu\rho}(q,0,-q) \,.
\label{Wproj}
\ee

The implementation of \1eq{Gammamu} 
at the level of the three ghost-gluon vertices
entering in the two diagrams of Fig.~\ref{fig:H_truncated} reveals that only the form factors $B_1$ survive the projection in \1eq{Wproj}. Therefore, the function $B_2$ does {\it not} contribute to ${\cal W}(q^2)$.

In order, to reduce  the complexity of the problem, all $B_1$ are considered to be functions
of one kinematic variable, namely 
the momentum of the gluon entering into the corresponding ghost-gluon vertex.
Specifically, introducing \mbox{$z := \ell - q $}, this approximation amounts to the replacements
\be
B_1(-z,\ell,-q)  \to   B_1(q^2)\,,\,\,  
B_1(-\ell,0,\ell)  \to   B_1(\ell^2)\,,\,\,   
B_1 (-z,0,z) \to {\overline B}_1(z^2,\ell^2)\,, 
\label{soft_aprox}
\ee
where ${\overline B}_1(z^2,\ell^2) := \frac{1}{2}[ B_1(z^2) +B_1(\ell^2)]$, originating from 
the implementation of the symmetrization procedure discussed in Sec.~III of~\cite{Aguilar:2018csq}.

The fully-dressed three-gluon vertex, 
\mbox{$\Gamma_{\mu\alpha\beta}(-q,-z,\ell)$}, appearing in diagram $(d_2)$, is approximated
by its tree-level structure multiplied by a unique form factor, $V$, depending only on $\ell^2$, \ie
\begin{align}
\Gamma_{\mu\alpha\beta}(-q,-z,\ell)=  V(\ell^2)\,\Gamma^{(0)}_{\mu\alpha\beta}(-q,-z,\ell) \,. 
\label{eq:3g}
\end{align}

With the aforementioned approximations, $K_{\nu\mu\rho}(q,0,-q)$ is given by
\be 
K_{\nu\mu\rho}(q,0,-q) =
\frac{1}{2} C_\mathrm{A} g^2 Z_1^\asym \left[ K^{d_1}_{\nu\mu\rho}(q,0,-q) + K^{d_2}_{\nu\mu\rho}(q,0,-q) \right] \,,
\label{HKp0}
\ee
where $Z_1^\asym$ is the renormalization constant given in \1eq{zasymnum},  
\bea
K^{d_1}_{\nu\mu\rho}(q,0,-q) &=& \int_\ell \Delta_{\rho\nu}(\ell)D(\ell^2)D(z^2)B_1(q^2) B_1(\ell^2)(q - \ell)_\mu \,, \nonumber\\
K^{d_2}_{\nu\mu\rho}(q,0,-q) &=& \int_\ell \Delta^\beta_{\nu}(\ell)\Delta^\alpha_\rho(z)D(z^2){\overline B}_1(z^2,\ell^2) V(\ell^2)\,\Gamma^{(0)}_{\mu\alpha\beta}(-q,-z,\ell) \,, 
\label{HKdiags}
\eea
and we resort to the compact notation for the integral over all space
\be 
\int_\ell := \frac{1}{( 2\pi )^4} \int d^4 \ell \,.
\ee

Projecting \1eq{HKdiags} as prescribed in \1eq{Wproj} leads to
\be 
{\cal W}(q^2) =
\frac{1}{6} C_\mathrm{A} g^2 Z_1^\asym \left[ {\cal W}_{d_1}(q^2) + {\cal W}_{d_2}(q^2) \right] \,,
\label{Wdiags}
\ee
where  
\bea
{\cal W}_{d_1}(q^2) &=& - i \int_\ell \frac{ \Delta(\ell^2)F(\ell^2)F(z^2)B_1(q^2) B_1(\ell^2)}{ q^2 \ell^4 z^2 } (\ell \cdot q )[ \ell^2 q^2 - ( \ell \cdot q )^2] \,, \nonumber\\
{\cal W}_{d_2}(q^2) &=& 2 i \int_\ell \frac{\Delta(\ell^2)\Delta(z^2)F(z^2){\overline B}_1(z^2,\ell^2) V(\ell^2) }{ q^2 \ell^2 z^4 } [ z^2 ( \ell \cdot q ) + (\ell\cdot q )^2 +  2 q^2 \ell^2 ] \,.
\label{mwfinal}
\eea
Finally, we convert \1eq{Wdiags} to Euclidean space, employ spherical coordinates, following the rules and conventions
described in Section~V of \cite{Aguilar:2018csq}. Note, in particular, that 
two of the angular integrations may be carried out trivially, furnishing a factor of $4 \pi$; the remaining 
integration over $\phi$ (the angle between $q$ and $\ell$) must be evaluated numerically. The final result reads 
\be 
{\cal W}(q^2) =\frac{ C_\mathrm{A} \alpha_s }{12\pi^2} \, Z_1^\asym \, q \left[ {\cal W}_{d_1}(q^2) + {\cal W}_{d_2}(q^2) \right] \,,
\label{a1tan_sde}
\ee
with  \mbox{$s_{\phi} := \sin \phi$}, \mbox{$c_{\phi} := \cos \phi$}, and 
\bea
\label{finalW}
{\cal W}_{d_1}(q^2)&=& B_1(q^2)\int_0^\infty\!\!\!\! d\ell^2 \ell \Delta(\ell^2) F(\ell^2) B_1(\ell^2) \int_0^\pi \!\!\! d\phi\, s^4_{\phi} c_{\phi}
\,\frac{F(z^2)}{z^2} \,,  
\\
{\cal W}_{d_2}(q^2) &=& - 2 \int_0^\infty\!\!\!\!  d\ell^2 \ell^3 \Delta(\ell^2)V(\ell^2)\!\int_0^\pi \!\!\!\! d\phi \,
s^4_{\phi} (z^2 c_\phi - \ell q s_\phi^2 + 3 \,\ell q ) \Delta(z^2) {\overline B}_1(z^2,\ell^2) \, \frac{F(z^2)}{z^4} \,. \nonumber
\eea
  
From \1eq{a1tan_sde} follows immediately that \mbox{${\cal W}(0) = 0$}.
   To estimate at what rate this limit is approached, we  
   resort to a one-loop ``massive''  calculation,  where the infrared finite propagator
   is replaced by the naive massive \mbox{$\Delta^{-1}(q^2)=q^2-m^2$},  
   while all other quantities are kept at their tree level values, \mbox{$F=B_1={\overline B}_1=V=1$}.
   Then, employing standard integration formulas at the level of
Eq.~\eqref{mwfinal}, we find that \mbox{$\lim\limits_{q^2\to 0}{\cal W}(q^2) \sim q^2 \ln(q^2/m^2)$}.

\acknowledgments 
The research of J.~P. is supported by the 
Spanish Ministry of Economy and Competitiveness (MINECO) under grant FPA2017-84543-P,
and the  grant  Prometeo/2019/087 of the Generalitat Valenciana. 
The work of  A.~C.~A. and M.~N.~F. are supported by the Brazilian National Council for Scientific and Technological Development (CNPq) under the grants 307854/2019-1,   142226/2016-5, and 464898/2014-5 (INCT-FNA). A.~C.~A. also acknowledges the financial support from  the
S\~{a}o Paulo Research Foundation (FAPESP) through the project 2017/05685-2.


%

\end{document}